\documentclass{revtex4}
\usepackage{graphicx}
\usepackage{epsfig}
\usepackage{psfrag}

\newcommand{\De}{\ensuremath{{\rm D}}}

\newcommand{\He}{\ensuremath{\null^4{\rm He}}}

\newcommand{\Li}{\ensuremath{\null^7{\rm Li}}}

\begin{document}

\title{\bf Early Universe Constraints on Time Variation of Fundamental
  Constants} 

\author{Susana J. Landau}
\affiliation{Departamento de F{í}sica, FCEyN, Universidad de
Buenos Aires, Ciudad Universitaria - Pab. 1, 1428 Buenos Aires,
Argentina}
\author{Mercedes E. Mosquera}
\affiliation{Facultad de Ciencias Astron\'{o}micas y
Geof\'{\i}sicas. Universidad Nacional de La Plata. Paseo del Bosque
S/N 1900 La Plata, Argentina}
\author{Claudia G. Scóccola}
\affiliation{Facultad de Ciencias Astron\'{o}micas y
Geof\'{\i}sicas. Universidad Nacional de La Plata. Paseo del Bosque
S/N 1900 La Plata, Argentina} \affiliation{Instituto de
Astrof{í}sica La Plata}
\author{Hector Vucetich}
\affiliation{Facultad de Ciencias Astron\'{o}micas y
Geof\'{\i}sicas. Universidad Nacional de La Plata. Paseo del Bosque
S/N 1900 La Plata, Argentina      }


\begin{abstract}
We study the time variation of fundamental constants in the early
Universe. Using data from primordial light nuclei abundances, CMB
and the 2dFGRS power spectrum, we put constraints on the time
variation of the fine structure constant $\alpha$, and the Higgs
vacuum expectation value $<v>$ without assuming any theoretical
framework. A variation in $<v>$ leads to a variation in the electron
mass, among other effects. Along the same line, we study the
variation of $\alpha$ and the electron mass $m_e$. In a purely
phenomenological fashion, we derive a relationship between both
variations.
\end{abstract}

\maketitle
\section{Introduction}
\label{Intro}

Unification theories, such as super-string
\cite{Wu86,Maeda88,Barr88,DP94,DPV2002a,DPV2002b}, brane-world
\cite{Youm2001a,Youm2001b,branes03a,branes03b} and Kaluza-Klein
theories \cite{Kaluza,Klein,Weinberg83,GT85,OW97}, allow fundamental
constants, such as the fine structure constant $\alpha$ and the
Higgs vacuum expectation value $<v>$, to vary over cosmological
timescales. A variation in $<v>$ leads to a variation in the
electron mass, among other effects. On the other hand, theoretical
frameworks based in first principles, were developed by different
authors \cite{Bekenstein82,BSM02,BM05} in order to study the
variation of certain fundamental constants. Since each theory
predicts a specific time behaviour, by setting limits on the time
variation of fundamental constants some of these theories could be
set aside.

Limits on the present rate of variation of $\alpha$ and
$\mu=\frac{m_e}{m_p}$ (where $m_e$ is the electron mass and $m_p$
the proton mass) are provided by atomics clocks
\cite{Bize03,Fischer04,Peik04,PTM95,Sortais00,Marion03}. Data from
the Oklo natural fission reactor \cite{Fujii00,DD96} and half lives
of long lived $\beta$ decayers \cite{Olive04b} allow to constrain
the variation of fundamental constants at $z \simeq 1$. Recent
astronomical data based on the analysis of spectra from
high-redshift quasar absorption systems suggest a possible variation
of $\alpha$ and $\mu$
\cite{Webb99,Webb01,Murphy01a,Murphy01b,Murphy03b,Ivanchik05,Tzana07}.
However, another analysis of similar data gives null variation of
$\alpha$ \cite{MVB04,QRL04,Bahcall04,Srianand04}.  Big Bang
Nucleosynthesis (BBN) and Cosmic Microwave Background (CMB) also
provide constraints on the variation of fundamental constants.
Although the limits imposed by BBN and CMB, are less stringent than
the previous ones, they are still important since they refer to the
earliest cosmological times.

In previous works, we have studied the time variation of the fine
structure constant in the early Universe to test Bekenstein model
\cite{Mosquera07} and the time variation of the electron mass to
test Barrow-Magueijo model \cite{Scoccola07}. However, unifying
theories predict relationships among the variation of gauge coupling
constants which depend on the theoretical framework. In this work,
we perform a phenomenological analysis of the joint time variation
of $\alpha$ and $<v>$ in the early Universe without assuming a
theoretical framework.

The model developed by Barrow \& Magueijo \cite{BM05} predicts the
variation of $m_e$ over cosmological timescales. This model could be
regarded as the low energy limit of a more sophisticated unified
theory. In such case, the unifying theory would also predict variation
of gauge coupling constants and in consequence the variation of
$\alpha$. Thus, in order to provide bounds to test such kind of models, we
also analyze in this paper the joint variation of $\alpha$ and $m_e$
without assuming a theoretical framework.


The dependence of the primordial abundances on $\alpha$ has been
analyzed by Bergstrom et al.\cite{Iguri99} and improved by Nollet \&
Lopez \cite{Nollet}, while the dependence on $<v>$ has been analyzed
by Yoo and Scherrer \cite{YS03}. Semi-analytical analyses have been
performed by some of us in earlier works \cite{LMV06,Chamoun07}.
Several authors \cite{CO95,Ichi02,ichi04} studied the effects of the
variation of fundamental constants on BBN in the context of a
dilaton superstring model.  M\"uuller et al \cite{mueller04}
calculated the primordial abundances as a function of the Planck
mass, fine structure constant, Higgs vacuum expectation value,
electron mass, nucleon decay time, deuterium binding energy, and
neutron-proton mass difference and studied the dependence of the
last three quantities as functions of the fundamental coupling and
masses. Coc et al. \cite{coc07} set constraints on the variation in
the neutron lifetime and neutron-proton mass difference using the
primordial abundance of $^4{\rm He}$. Cyburt et al. \cite{cyburt05}
studied the number of relativistic species at the time of BBN and
the variations in fundamental constants $\alpha$ and $G_N$. Dent et
al \cite{Dent07} studied the dependence of the primordial abundances
with nuclear physics parameters such as $G_N$, nucleon decay time,
$\alpha$, $m_e$, the average nucleon mass, the neutron-proton mass
difference and binding energies. Finally, limits on cosmological
variations of $\alpha$, $\Lambda_{QCD}$ and quark mass ($m^q$) from
optical quasar absorption spectra, laboratory atomic clocks and from
BBN have been established by Flambaum et al.
\cite{Flambaum02,Flambaum04b}.





In this paper, we study the effects of a possible variation of $\alpha$ and
$<v>$ on the primordial abundances, including the dependence of the
masses of the light elements on the cross sections, and using the
dependence on $<v>$ of the deuterium binding energy calculated in the
context of usual quantum theory with a phenomenological potential. We
use all available observational data of $\De$,$\He$ and $\Li$ to set
constraints on the joint variation of fundamental constants at the
time of BBN.

However, we do not consider a possible variation of $\Lambda_{QCD}$. Indeed,
the dependence of the physical quantities involved in the calculation
of the primordial abundances with a varying $\Lambda_{QCD}$ is highly
dependent on the model. The analyses of refs. \cite{CO95,Ichi02}, for
example, are done in the context of a string dilaton model. Therefore,
we will not consider such dependencies even though it has been analyzed
in the literature
\cite{CO95,KM03,Flambaum02,Flambaum04b,Ichi02}.  Our
  analysis, instead, is a model independent one.

Previous analysis of CMB data (earlier than the WMAP three-year
release) including a possible variation of $\alpha$ have been
performed by refs. \cite{Martins02,Rocha03,ichi06} and including a
possible variation of $m_e$ have been performed by
refs. \cite{YS03,ichi06}.  The work of Ichikawa et al. \cite{ichi06} is
the only one that assumes that both variations are related in the
context of string dilaton models.  In this work, we follow a
completely different approach, by assuming that the fundamental
constants vary independently.


The paper is organized as follows. In section \ref{nucleo}, we
present bounds on the variation of the fine structure constant and
the Higgs vacuum expectation value during Big Bang Nucleosynthesis.
We also discuss the difference between considering $<v>$ variation
and $m_e$ variation during this epoch. In section \ref{cmb}, we use
data from the CMB and from the 2dFGRS power spectrum to put bounds
on the variation $\alpha$ and $<v>$ (or $m_e$) during recombination,
allowing also other cosmological parameters to vary. In section
\ref{discusion}, we use the $\alpha-m_e$ and $\alpha-<v>$
confidence contours to obtain a phenomenological relationship
between both variations and then discuss our results. Conclusions are
presented in section \ref{conclusion}.

\section{Bounds from BBN}
\label{nucleo}

Big Bang Nucleosynthesis (BBN) is one of the most important tools to
study the early universe. The standard model has a single free
parameter, the baryon to photon ratio $\eta_B$, which can be
determined by comparison between theoretical calculations and
observations of the abundances of light elements. Independently, the
value of the baryonic density $\Omega_B h^2$ (related to $\eta_B$)
can be obtained with great accuracy from the analysis of the Cosmic
Microwave Background data \cite{wmapest,wmap3,Sanchez06}. Provided
this value, the theoretical abundances are highly consistent with
the observed ${\rm D}$ but not with all $^4{\rm He}$ and $^7{\rm
Li}$ data. If the fundamental constants vary with time, this
discrepancy might be solved and we may have insight into new physics
beyond the minimal BBN model.

In this section, we use available data of ${\rm D}$, $^4{\rm He}$ and
$^7{\rm Li}$ to put bounds on the joint variation of $\alpha$ and
$<v>$ and on the joint variation of $\alpha$ and $m_e$ at the time of
primordial nucleosynthesis.  The observational data for ${\rm D}$ have
been taken from refs.
\cite{pettini,omeara,kirkman,burles1,burles2,Crighton04,omeara06,oliveira06}.
For $^7{\rm Li}$ we consider the data reported by refs.
\cite{ryan,bonifacio1,bonifacio2,bonifacio3,Asplund05,BNS05,bonifacio07}.
For $^4{\rm He}$, we use the data from refs. \cite{PL07,izotov07} (see
ref.\cite{Mosquera07} for details).

We checked the consistence of the each group of data following ref.
\cite{PDGBook} and found that the ideogram method plots are not
Gaussian-like, suggesting the existence of unmodelled systematic
errors. We take them into account by increasing the errors by a
fixed factor, $2.10$, $1.40$ and $1.90$ for ${\rm D}$, $^4{\rm He}$
and $^7{\rm Li}$, respectively. A scaling of errors was also
suggested by ref. \cite{olive07}.

The main effects of the variation of the fine structure constant
during BBN are the variation of the neutron to proton ratio in thermal
equilibrium produced by a variation in the neutron-proton mass
difference, the weak decay rates and the cross sections of the
reactions involved during the first three minutes of the Universe. The
main effects of the variation of the Higgs vacuum expectation value
during BBN are the variation of the electron mass, the Fermi constant,
the neutron-proton mass difference and the deuterium binding energy,
affecting mostly the neutron to proton ratio, the weak decay rates and
the initial abundance of deuterium. In appendix \ref{correcciones} we
give more details about how the physics at BBN is modified by a
possible change in $\alpha$, $<v>$ and $m_e$. We modify the Kawano
code \cite{Kawano92} in order to consider time variation of $\alpha$
and $<v>$ and time variation of $\alpha$ and $m_e$ during BBN. The
coulomb, radiative and finite temperature corrections were included
following ref. \cite{Dicus82}.  We follow the analysis of
refs. \cite{Iguri99,Nollet} to introduce the variation in $\alpha$ on
the reaction rates.  The main effects of a change in $\alpha$ in
nuclear reaction rates are variations in the Coulomb barrier for
charged-induced reactions and radiative captures. We introduce the
dependence of the light nuclei masses on $\alpha$, correction that
affect the reaction rates, their inverse coefficients and their
Q-values \cite{LMV06}. We also update the value of the reaction rates
following ref. \cite{Iguri99}.

\begin{figure}[!ht]
\begin{center}
\includegraphics[scale=0.40,angle=-90]{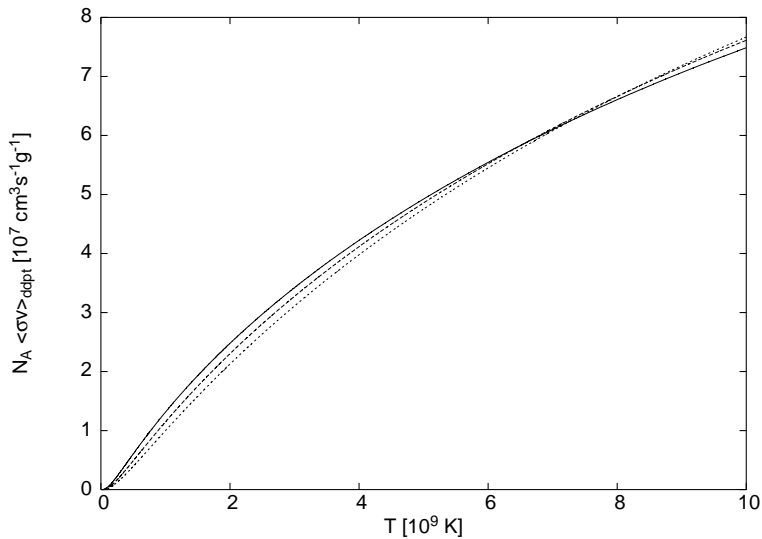}
\end{center}
\caption{$N_A <\sigma v>$ (in units of $10^7$ cm$^3$ s$^{-1}$
g$^{-1}$) for the reaction $\rm d + d \to p + t$, as a
function of the temperature (in units of $10^9$ K), when
$\frac{\Delta \alpha}{\alpha_0}=-0.1$ (solid line), $\frac{\Delta
\alpha}{\alpha_0}=0.0$ (dashed line) and $\frac{\Delta
\alpha}{\alpha_0}=0.1$ (dotted line)} \label{ddpt-al}
\end{figure}

To illustrate the effect of the variation in the fine structure
constant on the reactions rates, we present  the nuclear reaction
rate of $\rm d + d \to p + t$ $\left(R[dd;pt]=0.93 ×
10^{-3} \Omega_B h^2 T_9^3 N_A <\sigma v> \right)$ as a function of
$\frac{\Delta \alpha}{\alpha_0}$ ($\Delta \alpha=\alpha - \alpha_0$
and $\alpha_0$ is the current value of the fine structure constant):
\begin{eqnarray}
R[dd;pt]&=&2.369 × 10^{-3} \Omega_B h^2 T_9^{7/3}
{\alpha_0}^{1/3} \left(1+ \frac{\Delta
\alpha}{\alpha_0}\right)^{4/3}  {\mu}^{-1/3} e^{-9.545×
10^{10}\left(\frac{\mu \alpha_0^2}{T_9}\left(1+ \frac{\Delta
\alpha}{\alpha_0}\right)^2 \right)^{1/3}} \left[
1+0.16 \left(1+ \frac{\Delta \alpha}{\alpha_0}\right) \right] × \nonumber \\
 && \left(1+4.365 × 10^{-12}{\mu}^{-1/3} {\alpha_0}^{-2/3} \left(1+ \frac{\Delta \alpha}{\alpha_0}\right)^{-2/3}
 T_9^{1/3}+1.161 × 10^{10}  {\mu}^{1/3}
  {\alpha_0}^{2/3} \left(1+ \frac{\Delta \alpha}{\alpha_0}\right)^{2/3} T_9^{2/3}\right. \\
 &&  \hskip 0.5cm \left.+0.355 T_9-5.104 × 10^{18} {\mu}^{2/3} {\alpha_0}^{4/3} \left(1+ \frac{\Delta \alpha}{\alpha_0}\right)^{4/3}
  T_9^{4/3}-3.966 × 10^{8}{\mu}^{1/3}
 {\alpha_0}^{2/3} \left(1+ \frac{\Delta \alpha}{\alpha_0}\right)^{2/3} T_9^{5/3}\right)\, \, ,\nonumber
\end{eqnarray}
where $T_9$ is the temperature in units of $10^9$ K and $\mu$ is the
reduced mass. The reduced mass also changes if the fine structure
constant varies with time (see appendix \ref{correcciones}). This
nuclear reaction is important for calculating the final deuterium
abundance since this reaction destroys deuterium and produces
tritium which is crucial to form $^4{\rm He}$. In Figure
\ref{ddpt-al} we present the value of $N_A <\sigma v>$ for this
reaction as a function of the temperature, for different values of
$\frac{\Delta \alpha}{\alpha_0}$. If the fine structure constant is
greater than its present value, the reaction rate is lower than in
the case of no $\alpha$ variation. A decrease in the value of this
reaction rate results in an increase in the deuterium abundance.

The $^4{\rm He}$ abundance is less sensitive to changes in the nuclear
reaction rates than the other abundances (deuterium and $^7{\rm Li}$)
\cite{Iguri99} and very sensitive to variations in the parameters that
fixed the neutron-to-proton ratio. In thermal equilibrium, this ratio
is:
\begin{eqnarray}
\frac{Y_n}{Y_p}&=&e^{-\Delta m_{np}/T}\, \, ,
\end{eqnarray}
where $Y_n$ $\left(Y_p\right)$ is the neutron (proton) abundance,
$\Delta m_{np}$ is the neutron-proton mass difference and $T$ is the
temperature in MeV. When the weak interaction rates become slower
than the Universe expansion rate the neutron-to-proton ratio
freezes-out at temperature $T_f$. Afterwards, nearly all the
available neutrons are captured in $^4{\rm He}$ \cite{Iguri99}, this
abundance can be estimated by:
\begin{eqnarray}
Y_4&\sim& 2 \left(\frac{Y_n}{Y_p}\right)_f
\left[1+\left(\frac{Y_n}{Y_p}\right)_f\right]^{-1}\,\, ,
\end{eqnarray}
where $\left(\frac{Y_n}{Y_p}\right)_f=e^{-\Delta m_{np}/T_f}$. The
neutron-proton mass difference is affected by a change in the fine
structure constant and in the Higgs vacuum expectation value:
\begin{eqnarray}
\frac{\delta \Delta m_{np}}{\Delta m_{np}} &=& -0.587 \frac{\Delta
\alpha}{\alpha_0}+ 1.587 \frac{\Delta <v>}{<v>_0}\, \, .
\end{eqnarray}
An increase in the fine structure constant results in a decrease in
$\Delta m_{np}$, this produces a larger equilibrium
neutron-to-proton ratio and a larger abundance of $^4{\rm He}$.
However, an increase in $<v>$ leads to an increase in $\Delta
m_{np}$. This produces a smaller equilibrium neutron-to-proton
equilibrium ratio and a smaller abundance of $^4{\rm He}$
\cite{YS03}.

The freeze-out temperature of weak interactions is crucial to
determinate the amount of  available neutrons and therefore the
primordial abundance of $^4{\rm He}$. This temperature is modified
if the Higgs vacuum expectation value is changed during BBN due to
changes in the weak reaction rates (see appendix
\ref{correcciones}). A larger Higgs vacuum expectation value during
BBN results in: i) a smaller $G_F$ leading to earlier freeze-out of
the weak reactions $\left(n\leftrightarrow p \right)$, producing
more $^4{\rm He}$; ii) an increase in $m_e$, a decreasing of
$n\leftrightarrow p $ reaction rates and also producing more $^4{\rm
He}$ \cite{YS03}.

The dependence of the deuterium binding energy on the Higgs vacuum
expectation value is extremely model dependent. Bean and Savage
\cite{BS03b} studied this dependence using chiral perturbation
theory and their results were applied by several authors
\cite{YS03,mueller04,KM03}. We performed another estimation in the
context of usual quantum theory, using the effective Reid potential
\cite{reid68} for the nucleon-nucleon interaction (paper in
preparation).  Even though Yoo and Scherrer had shown that very
different values for $\frac{\partial \epsilon_D}{\partial m_\pi}$
lead to similar constraints on the change of $<v>$, we perform our
calculation using two different relationships (see Table \ref{ed}):
i) the obtained by ref. \citep{YS03} using the results of ref.
\cite{BS03b}; ii) the obtained using the effective Reid potential.
From Table \ref{ed}, it follows that the value obtained using the
effective Reid potential lies in the range allowed by the estimation
of \citet{BS03b}. The variation of the deuterium binding energy due
to a time variation of the Higgs vacuum expectation value is related
to $\frac{\partial \epsilon_D}{\partial m_\pi}$ as:
\begin{eqnarray}
\frac{\Delta \epsilon_D}{\left(\epsilon_D\right)_0}= \frac{\partial
\epsilon_D}{\partial m_\pi} \frac{m_\pi}{2\left(\epsilon_D\right)_0}
\frac{\Delta <v>}{<v>_0}\, \, .
\end{eqnarray}
We call $\kappa= \frac{m_\pi}{2\left(\epsilon_D\right)_0}
\frac{\partial \epsilon_D}{\partial m_\pi}$ hereafter.
\begin{table}[!ht]
\renewcommand{\arraystretch}{1.3}
\begin{center}
\caption{Values used in this work for $\frac{\partial
\epsilon_D}{\partial m_\pi}$ and the coefficient $\kappa$ in the
relationship $\frac{\Delta \epsilon_D}{\left(
\epsilon_D\right)_0}=\kappa \frac{\Delta <v>}{<v>_0}$.} \label{ed}
\begin{tabular}{|c|c|c|}
\hline & $\frac{\partial \epsilon_D}{\partial m_\pi}$&$\kappa$ \\
\hline Yoo and Scherrer & $-0.159$& $-5.000$\\\hline Reid potential &$-0.198$&$-6.230$ \\
\hline
\end{tabular}
\end{center}
\end{table}

An increase in the Higgs vacuum expectation value results in a
decrease in the deuterium binding energy, leading to an smaller
initial deuterium abundance:
\begin{eqnarray}
Y_d&=&\frac{Y_n Y_p \, \, e^{11.605 \epsilon_D/T_9}}{0.471×
10^{-10}T_9^{3/2}}\, \, ,
\end{eqnarray}
where $\epsilon_D$ is in MeV. The production of $^4{\rm He}$ begins later,
leading to a smaller helium abundance but also to an increase in the
final deuterium abundance \cite{YS03}.

To assume  time variation of the electron mass during BBN is not
exactly the same as to assume  time variation of the Higgs vacuum
expectation value since the weak interactions are important during
this epoch. There exist some well tested theoretical model that
predict time variation of the electron mass \cite{BM05}. For this
reason we compute the light nuclei abundances and perform a
statistical analysis using the observational data mentioned above to
obtain the best fit values for the parameters for the following
cases:
\begin{itemize}
\item variation of $\alpha$ and $<v>$ allowing $\eta_B$ to vary,
\item variation of $\alpha$ and $<v>$ keeping $\eta_B$ fixed,
\item variation of $\alpha$ and $m_e$ allowing $\eta_B$ to vary,
\item variation of $\alpha$ and $m_e$ keeping $\eta_B$ fixed.
\end{itemize}

Even though the WMAP data are able to constrain the baryon density
with great accuracy, there is still some degeneracy between the
parameters involved in the statistical analysis. For this reason, we
allow the joint variation of baryon density and the other two
constants to obtain an independent estimation for $\eta_B$. In the
cases were $\eta_B$ is fixed, we assume the value reported by the
WMAP team \cite{wmap3}  $\left(\eta_B^{WMAP}= (6.108 \pm 0.219)
× 10^{-10} \right)$. We also present in Table \ref{eta}, as an
independent estimation, the best fit for the baryon density when all
the constants are fixed at their present value. It is shown that the
only reasonable fit is found when the lithium data is removed from
the data set, and the value for $\eta_B$ is consistent with the
value reported by \citet{wmap3}. The fits considering the data of
$^7{\rm Li}$ are not reasonable and the best fit for $\eta_B$ is not
consistent with the WMAP value.

\begin{table}[h!]
\begin{center}
\caption{Best fit parameter values, 1$\sigma$ errors for the BBN
constraints on $\eta_B$ (in units of $10^{-10}$) keeping all the
fundamental constants fixed at their present value.} \label{eta}
\begin{tabular}{|c|c|c|}
\hline & $\eta_B \pm \sigma \left[ 10^{-10}\right]$&$\frac{\chi^2_{min}}{N-1}$ \\
\hline
 ${\rm D}+ ^4{\rm He}+ ^7{\rm Li}$&  $4.310 \pm 0.050$& $10.00$ \\ \hline
 $^4{\rm He}+ ^7{\rm Li}$  & $3.920 \pm 0.080$&  $6.53$ \\ \hline
 ${\rm D}+ ^7{\rm Li}$  & $4.270 \pm 0.060$&  $9.27$ \\ \hline
 ${\rm D}+ ^4{\rm He}$ & $6.710_{-0.360}^{+0.400}$ & $1.61$ \\ \hline
\end{tabular}
\end{center}
\end{table}

Table \ref{results-table1} shows the results for the analysis of the
variation of $\alpha$ and $<v>$ when $\eta_B$ is allowed to vary.
These results correspond to the relationship between $\epsilon_D$
and $<v>$ obtained using Reid potential. This fit is consistent
within 1$\sigma$ with the acquired considering the value of $\kappa$
calculated by Yoo and Scherrer. We will not perform the statistical
analysis again excluding one group of data for the case where
$\eta_B$ is allowed to vary since we would have two groups of data
and three unknown variables. A reasonable fit can be found when
$\alpha$, $<v>$ and $\eta_B$ are allowed to vary. This fit is
consistent within 6$\sigma$ with non null values for the variations
of $\alpha$ and $<v>$ while the obtained value for $\eta_B$ is not
consistent with the estimation of WMAP within 3$\sigma$.

\begin{figure}[!ht]
\begin{center}
\includegraphics[scale=0.40,angle=0]{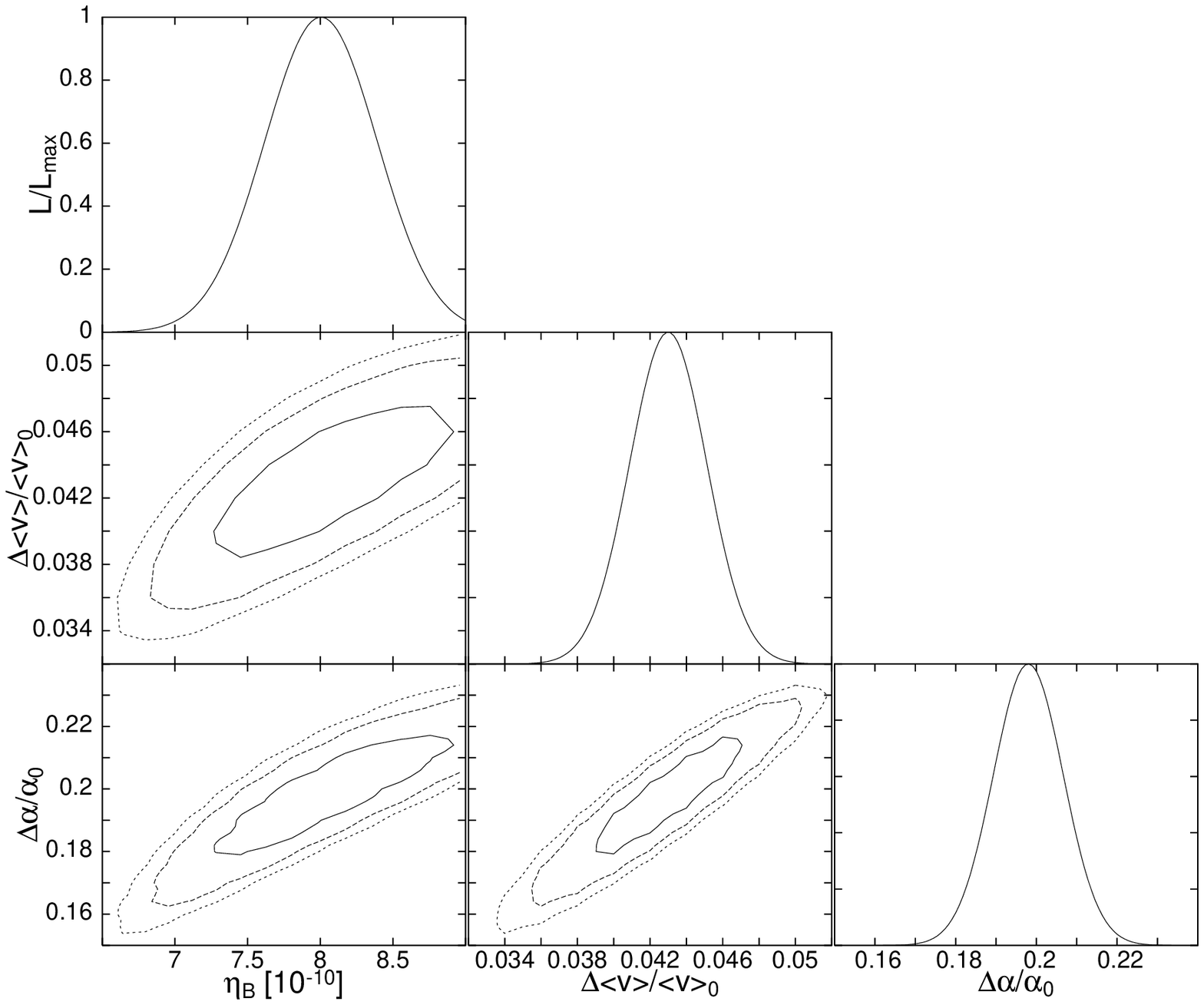}
\includegraphics[scale=0.40,angle=0]{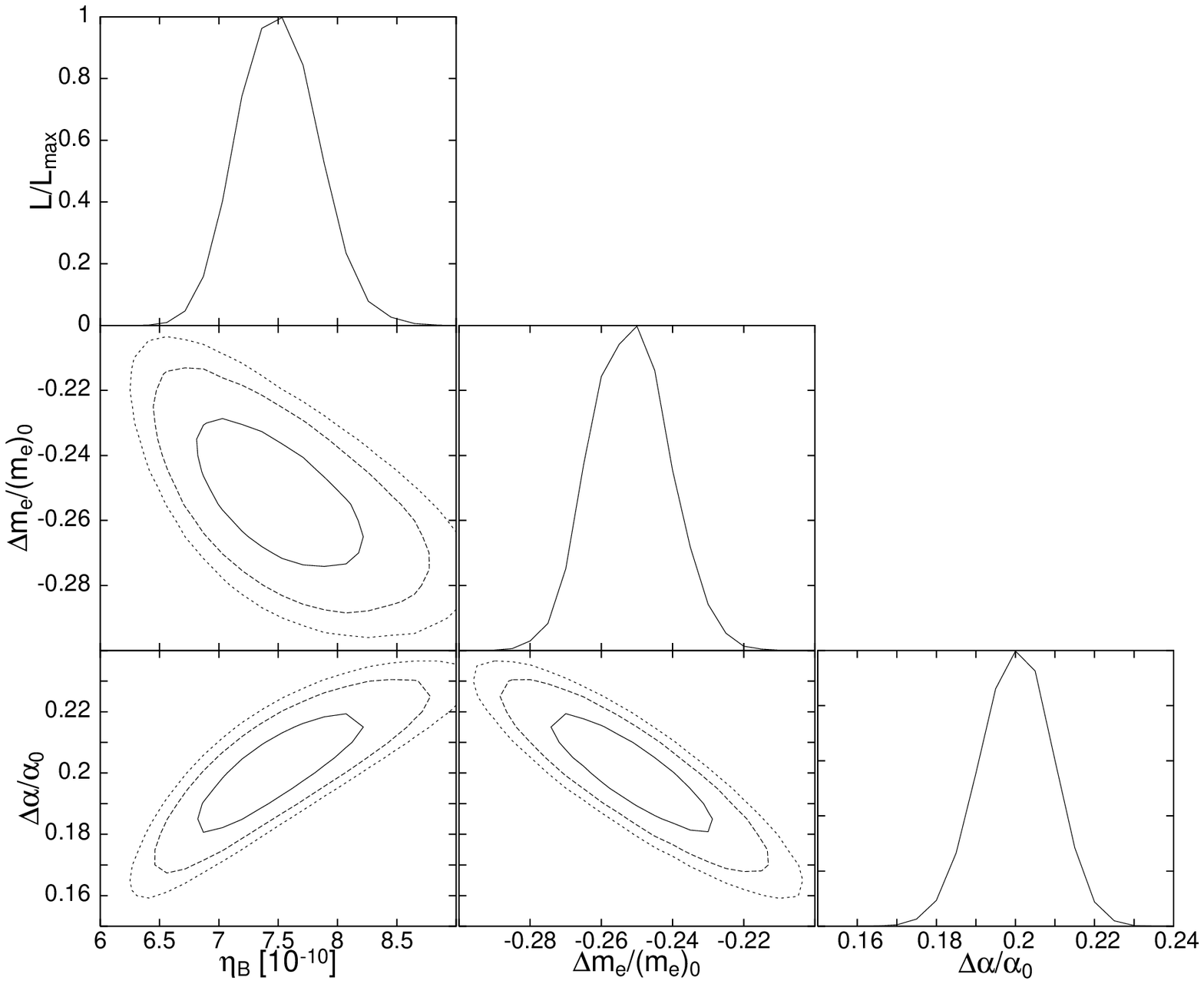}
\end{center}
\caption{Left Figure: Likelihood contours for $\frac{\Delta
\alpha}{\alpha_0}$, $\frac{\Delta <v>}{<v>_0}$ and $\eta_B$ (in
units of $10^{-10}$) and 1 dimensional Likelihood (using
$\frac{\Delta \epsilon_D}{\left( \epsilon_D\right)_0}=-6.230
\frac{\Delta <v>}{<v>_0}$). Right Figure: Likelihood contours for
$\frac{\Delta \alpha}{\alpha_0}$, $\frac{\Delta
m_e}{\left(m_e\right)_0}$ and $\eta_B$ (in units of $10^{-10}$) and
1 dimensional Likelihood.} \label{al-me-eta-fig}
\end{figure}

Left Figure \ref{al-me-eta-fig} shows a
strong degeneracy between $\frac{\Delta \alpha}{\alpha_0}$ and
$\eta_B$, $\frac{\Delta <v>}{<v>_0}$ and $\eta_B$ and
$\frac{\Delta\alpha}{\alpha_0}$ and $\frac{\Delta <v>}{<v>_0}$, ($\Delta
<v>=<v>-<v>_0$ and $<v>_0$ is the present value of the Higgs
vacuum expectation value). From this phenomenological approach, the
variation of the fundamental constants can be used to reconcile
the observed primordial abundances.

\begin{table}[!ht]
\renewcommand{\arraystretch}{1.3}
\begin{center}
\caption{Best fit parameter values and 1$\sigma$ errors for the BBN
constraints on $\frac{\Delta\alpha}{\alpha_0}$, $\frac{\Delta
<v>}{<v>_0}$,$\frac{\Delta m_e}{\left(m_e\right)_0}$, allowing
$\eta_B$ (in units of $10^{-10}$) to vary and considering ${\rm D}+
^4{\rm He}+ ^7{\rm Li}$. We use the estimation $\frac{\Delta
\epsilon_D}{\left( \epsilon_D\right)_0}=-6.230 \frac{\Delta
<v>}{<v>_0}$ to obtain the fit on $\frac{\Delta\alpha}{\alpha_0}$,
$\frac{\Delta <v>}{<v>_0}$ and $\eta_B$.} \label{results-table1}
\begin{tabular}{|c|c|c|c|c|}
\hline $\frac{\Delta \alpha}{\alpha_0}\pm \sigma$&$\frac{\Delta
<v>}{<v>_0}\pm \sigma$&$\frac{\Delta m_e}{\left(m_e\right)_0}\pm
\sigma$&$\eta_B \pm \sigma [10^{-10}]$&$\frac{\chi^2_{min}}{N-3}$ \\
\hline
$0.198^{+0.013}_{-0.014}$&$0.043^{+0.003}_{-0.004}$&---&$8.005^{+0.552}_{-0.553}$
&$1.16$ \\ \hline \hline
$0.210_{-0.013}^{+0.015}$&---&$-0.250_{-0.018}^{+0.015}$&$7.533_{-0.502}^{+0.447}$&1.11 \\
\hline
\end{tabular}
\end{center}
\end{table}

When $\alpha$, $m_e$ and $\eta_B$ are considered as free parameters,
we also obtain a reasonable fit (see Table \ref{results-table1}).
Once again, we will not perform the statistical analysis again
excluding one group of data since we would have two groups of data
and three unknown variables. There is consistency within 6$\sigma$
with variation of $\alpha$ and $m_e$ but the obtained value for
$\eta_B$ is not consistent with the estimation of WMAP within
3$\sigma$. Right Figure \ref{al-me-eta-fig} shows a strong
degeneracy between $\frac{\Delta \alpha}{\alpha_0}$ and $\eta_B$,
$\frac{\Delta m_e}{\left(m_e\right)_0}$ and $\eta_B$ and
$\frac{\Delta\alpha}{\alpha_0}$ and $\frac{\Delta
m_e}{\left(m_e\right)_0}$ $\left(\Delta
m_e=m_e-\left(m_e\right)_0\right.$, $\left(m_e\right)_0$ is the
present value of the electron mass). From this result it might be
possible to reconcile the observed primordial abundances and the
WMAP estimation for the baryon density.

By comparing the results presented in Table \ref{results-table1} and
in Table \ref{eta}, it can be noticed that the variation of the
fundamental constants improves the statistical analysis providing a
$\chi^2_{min}/\left(N-3\right)$ value that is closer to one.

\begin{figure}[!ht]
\begin{center}
\includegraphics[scale=0.55,angle=0]{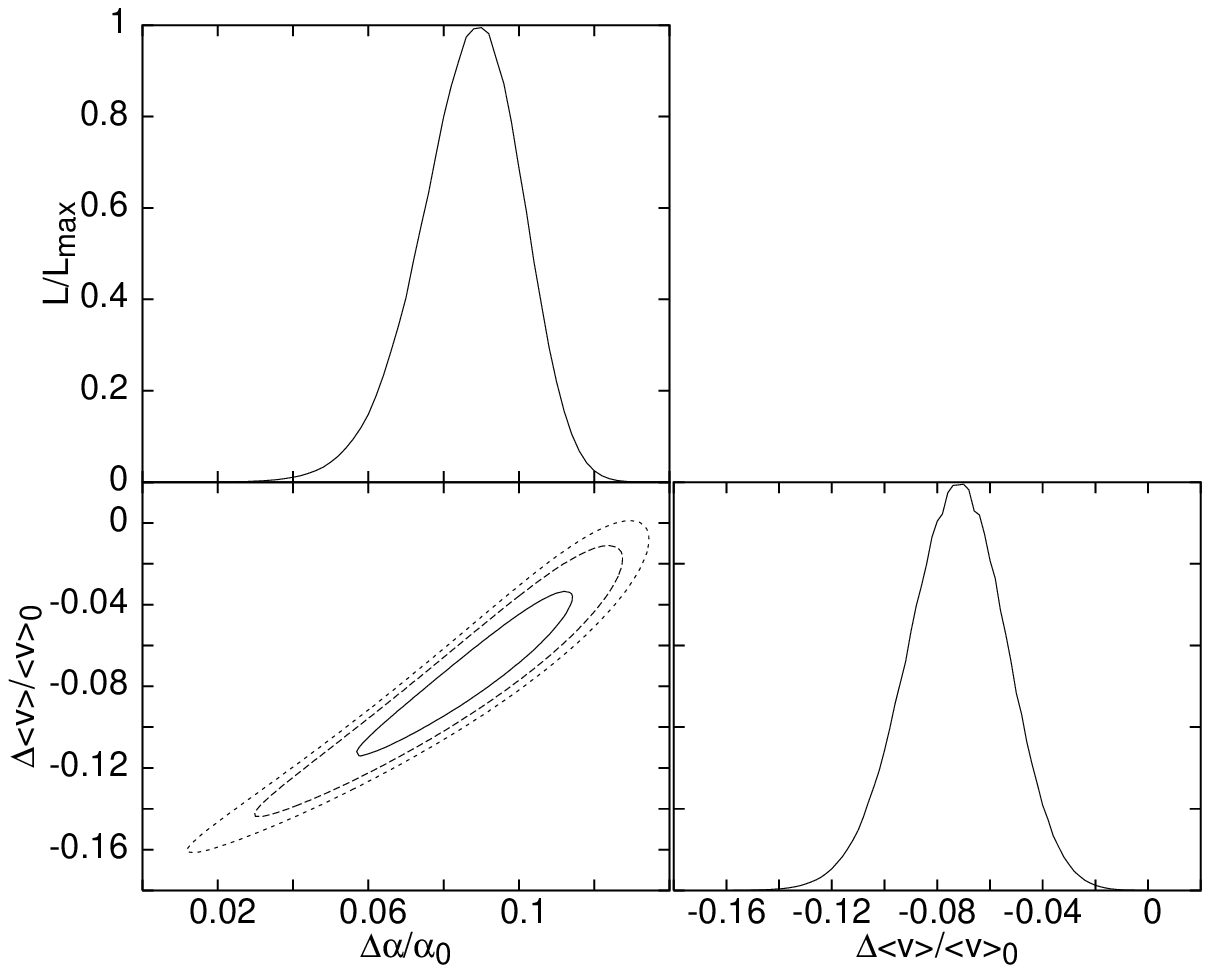}
\includegraphics[scale=0.55,angle=0]{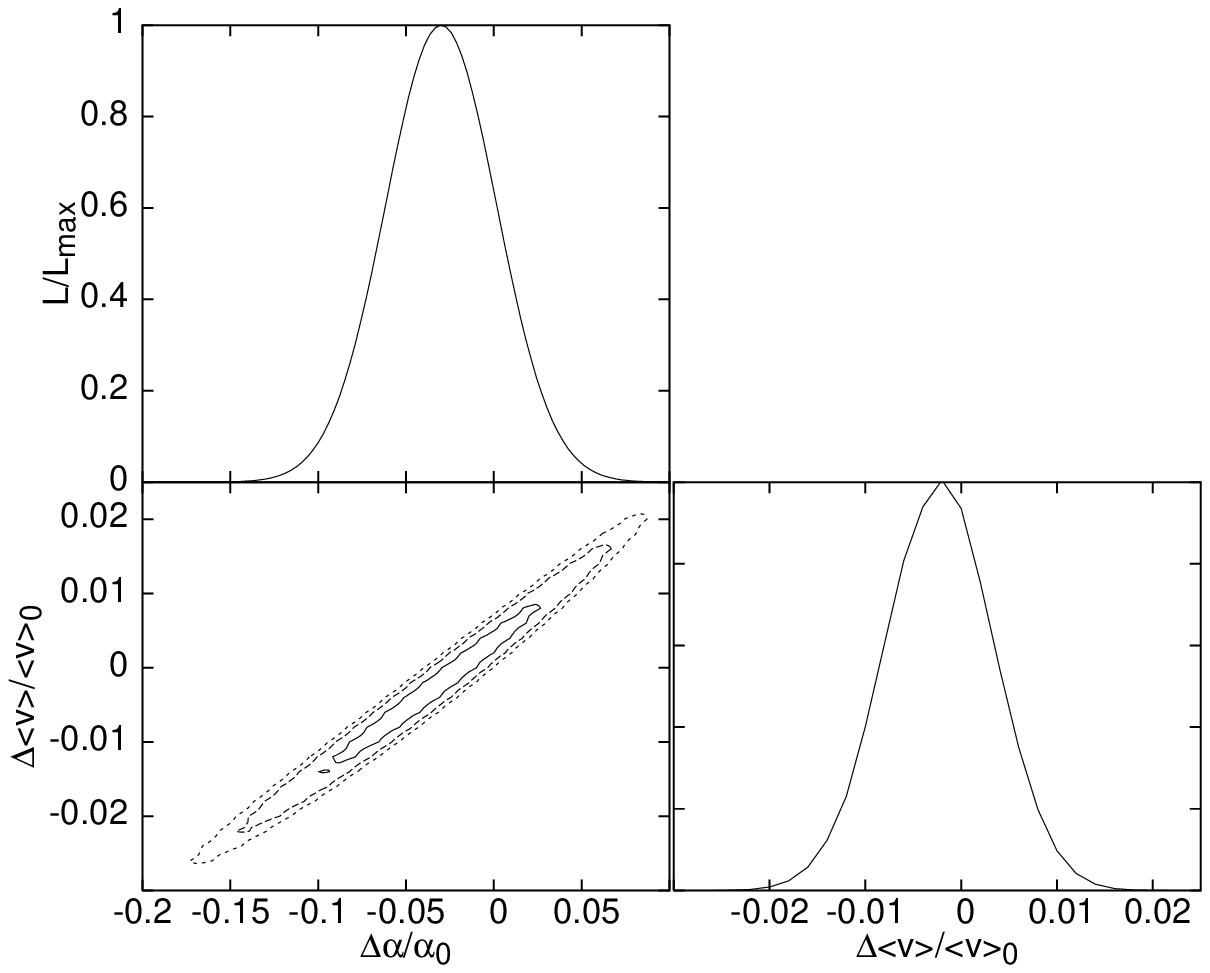}
\end{center}
\caption{Likelihood contours for $\frac{\Delta \alpha}{\alpha_0}$ vs
$\frac{\Delta  <v>}{<v>_0}$ and 1 dimensional Likelihood. Left
Figure: ${\rm D} + ^7{\rm Li}$ data; Right Figure: ${\rm D} +^4{\rm
He}$. The Figures were obtained using $\frac{\Delta
\epsilon_D}{\left( \epsilon_D\right)_0}=-6.230 \frac{\Delta
<v>}{<v>_0}$.} \label{al-v-fig}
\end{figure}

We now consider the joint variation of the fine structure constant
and the Higgs vacuum expectation value with $\eta_B$ fixed at the
WMAP estimation. In this case, it is reasonable to repeat the
analysis excluding one group of data at the time. The results are
presented in Table \ref{results-table-v} and were obtained using
$\frac{\Delta \epsilon_D}{\left( \epsilon_D\right)_0}=-6.230
\frac{\Delta <v>}{<v>_0}$. The fits obtained using $\frac{\Delta
\epsilon_D}{\left( \epsilon_D\right)_0}=-5.000 \frac{\Delta
<v>}{<v>_0}$ are consistent, within 1$\sigma$, with the ones
presented. There is good fit for the whole data set and also
excluding one group of data at each time. In any case, there is a
strong degeneracy between $\frac{\Delta \alpha}{\alpha_0}$ and
$\frac{\Delta <v>}{<v>_0}$ (see Figure \ref{al-v-fig}). Considering
all data or $^4{\rm He} + ^7{\rm Li}$ we find variation of both
fundamental constants, $\alpha$ and $<v>$, even at 6$\sigma$.
However, if the statistical analysis is performed with ${\rm
D}+^4{\rm He}$ we find null variation for both constants within
1$\sigma$. For completeness we also modified the Kawano's code in
order to calculate the different primordial abundances for two
values inside the range of $\frac{\partial \epsilon_D}{\partial
m_\pi}$ calculated by refs. \cite{EMG03,YS03} $\left( -0.15 <
\frac{\partial\epsilon_D}{\partial m_\pi} < -0.05\right)$ and
performed the statistical test in order to obtain the constraints on
the variation of the fundamental constants $\alpha$ and $<v>$. All
the results are consistent within 2$\sigma$ with the ones presented
above.
\begin{table}[!ht]
\renewcommand{\arraystretch}{1.3}
\begin{center}
\caption{Best fit parameter values, 1$\sigma$ errors for the BBN
constraints on $\frac{\Delta\alpha}{\alpha_0}$ and $\frac{\Delta
<v>}{<v>_0}$, with $\eta_B$ fixed at the WMAP estimation. The
results correspond to the estimation $\frac{\Delta
\epsilon_D}{\left( \epsilon_D\right)_0}=-6.230 \frac{\Delta
<v>}{<v>_0}$.} \label{results-table-v}
\begin{tabular}{|c|c|c|c|}
\hline Data &$\frac{\Delta \alpha}{\alpha_0}\pm \sigma$&
$\frac{\Delta <v>}{<v>_0}\pm \sigma$&$\frac{\chi^2_{min}}{N-2}$\\
\hline ${\rm D}+ ^4{\rm He}+ ^7{\rm Li}$& $0.140 \pm 0.006$&$0.032
\pm 0.002 $ &$2.52$\\ \hline $^4{\rm He}+ ^7{\rm Li}$&
$0.148_{-0.008}^{+0.004}$&$0.033_{-0.003}^{+0.002}$&$1.23$ \\
\hline ${\rm D}+^7{\rm Li}$&
$0.090_{-0.022}^{+0.017}$&$-0.070_{-0.026}^{+0.024}$&$1.15$\\ \hline
${\rm D}+ ^4{\rm He}$&
$-0.030_{-0.030}^{+0.035}$&$-0.002_{-0.008}^{+0.007}$&$1.03$
\\ \hline
\end{tabular}
\end{center}
\end{table}

\begin{figure}[!ht]
\begin{center}
\includegraphics[scale=0.55,angle=0]{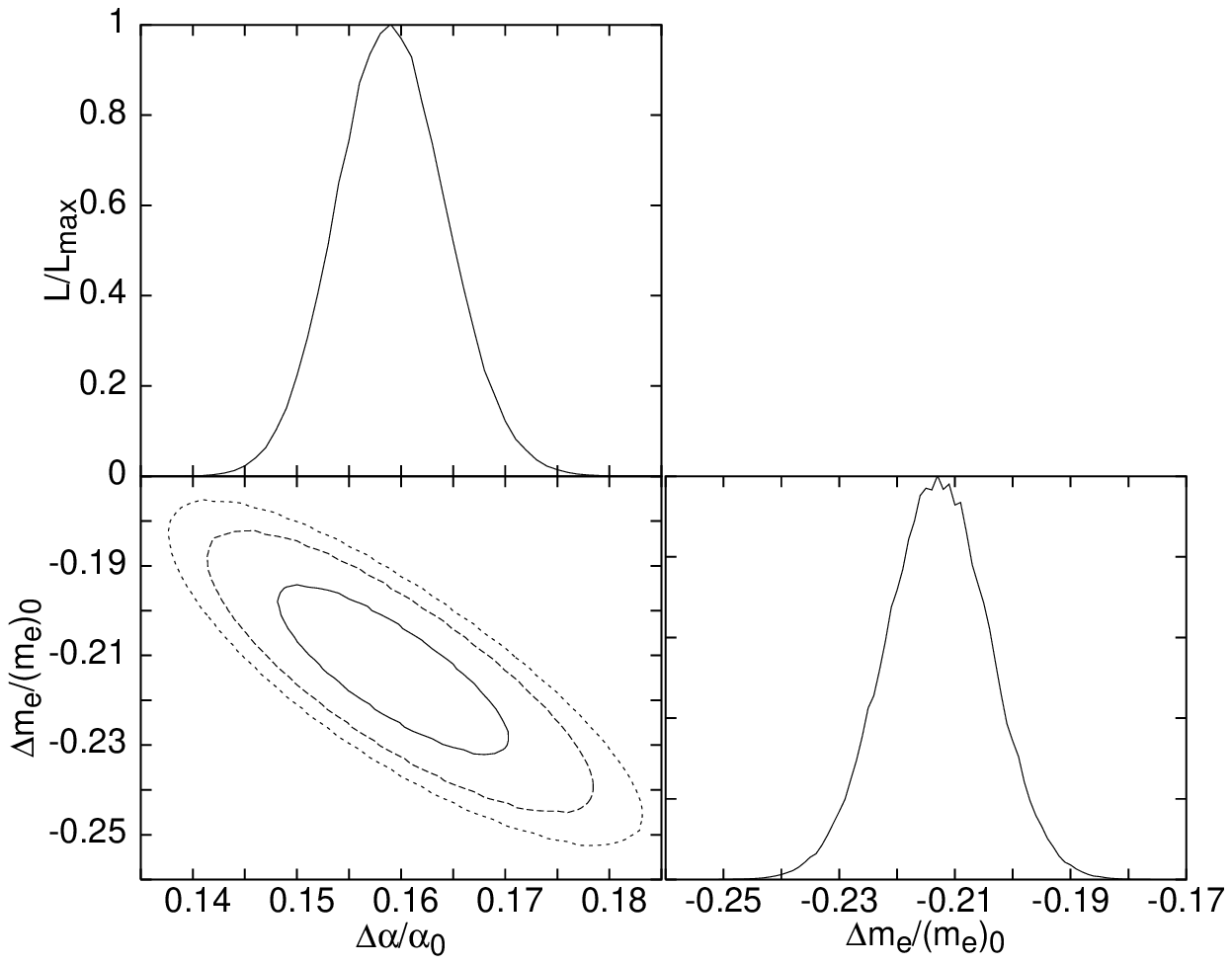}
\includegraphics[scale=0.55,angle=0]{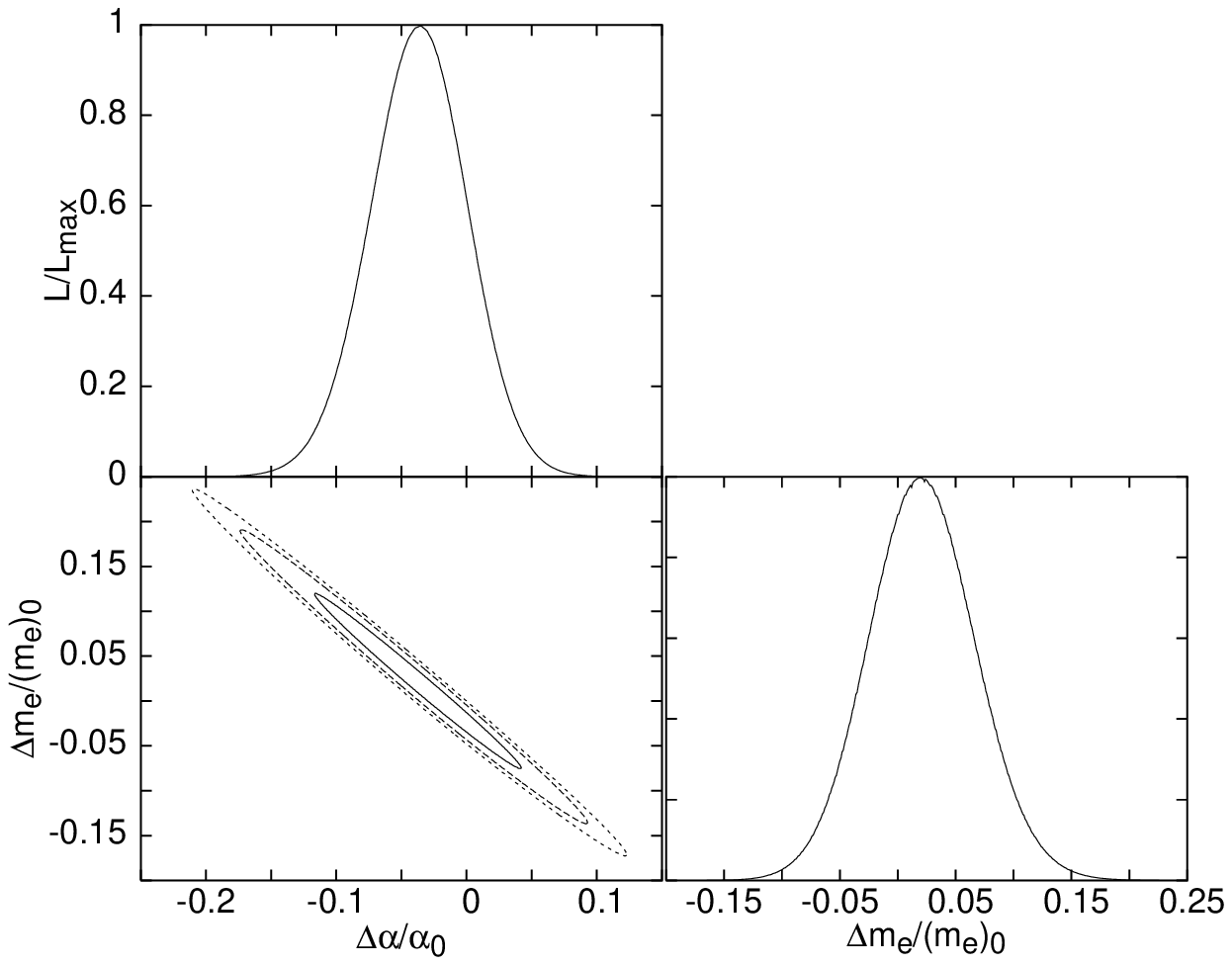}
\end{center}
\caption{Likelihood contours for $\frac{\Delta \alpha}{\alpha_0}$ vs
$\frac{\Delta m_e}{\left(m_e\right)_0}$ and 1 dimensional
Likelihood. Left Figure: ${\rm D}+ ^4{\rm He}+ ^7{\rm Li}$ data;
Right Figure: ${\rm D}+ ^4{\rm He}$.} \label{al-me-fig1}
\end{figure}

Table \ref{results-table-me} and Figure \ref{al-me-fig1} show the
results obtained when only $\alpha$ and $m_e$ are allowed to vary.
There is a strong degeneracy between the variations of $\alpha$ and
the variations of $m_e$ in all the cases considered. We find
reasonable fits for the whole data set and also excluding one group
of data at each time. Considering all data or $^4{\rm He}+^7{\rm
Li}$, we find variation of $\alpha$ and $m_e$, even at 6$\sigma$. On
the other hand, if the statistical analysis is performed with ${\rm
D}+^4{\rm He}$ we find null variation for both constants within
1$\sigma$.

\begin{table}[!ht]
\renewcommand{\arraystretch}{1.3}
\begin{center}
\caption{Best fit parameter values, 1$\sigma$ errors for the BBN
constraints on $\frac{\Delta\alpha}{\alpha_0}$ and $\frac{\Delta
m_e}{\left(m_e\right)_0}$, with $\eta_B$ fixed at the WMAP estimation.} \label{results-table-me}
\begin{tabular}{|c|c|c|c|c|}
\hline Data &$\frac{\Delta \alpha}{\alpha_0}\pm \sigma$&
$\frac{\Delta m_e}{\left(m_e\right)_0}\pm \sigma $&$\frac{\chi^2_{min}}{N-2}$  \\
\hline ${\rm D}+ ^4{\rm He}+ ^7{\rm Li}$& $0.159\pm
0.008$&$-0.213\pm 0.012$&$1.85$
\\ \hline
$^4{\rm He}+ ^7{\rm Li}$ & $0.163\pm 0.008$&$-0.218\pm 0.013$&$1.00$
\\ \hline
${\rm D}+ ^7{\rm Li}$ & $0.067_{-0.015}^{+0.022}$&$0.447 \pm
0.134$&$1.00$
\\ \hline
${\rm D}+ ^4{\rm He}$ &
$-0.036_{-0.053}^{+0.052}$&$0.020_{-0.064}^{+0.066}$&$1.00$\\
\hline
\end{tabular}
\end{center}
\end{table}

Richard et al. \cite{richard05} have pointed out that a better
understanding of turbulent transport in the radiative zones of the
stars is needed in order to get a reliable estimation of the $^7{\rm
Li}$ abundance, while Menendez and Ramirez \cite{MR04} have
reanalyzed the $^7{\rm Li}$ data and obtained results that are
marginally consistent with the WMAP estimate. On the other hand,
Prodanovic and Fields \cite{PF07} put forward that the discrepancy
with the WMAP data can worsen if contamination with $^6{\rm Li}$ is
considered. Therefore, we adopt the conservative criterion that the
bounds on the variation of fundamental constants obtained in this
paper are those where only the data of ${\rm D}$ and $^4{\rm He}$
are fitted to the theoretical predictions of the abundances. This
paper shows evidence for variation of fundamental constants in the
early Universe if the reported values for the $\Li$ abundance are
confirmed by future observations and/or improvement of the
theoretical analyses.

In Table \ref{resumenbbn} we summarize our results for the variation
of $\alpha$, $<v>$ and $m_e$, using ${\rm D}+^4{\rm He}$ data in the
statistical analysis. The sign of  $\alpha$ variation is the same
for all the cases.  However, the sign of the variation of $<v>$ or
$m_e$ changes depending on whether the joint variation with $\alpha$
is considered or not. In all cases, we have found null variation of
the fundamental constants within 3$\sigma$ and the values of
$\frac{\chi^2_{min}}{N-g}$ (where $g=2$ for the case where
 two constants are allowed to vary, and $g=1$ when only one
fundamental constant is allowed to vary) are closer to one,
resulting in reasonable fits for all the cases.
\begin{table}[!ht]
\renewcommand{\arraystretch}{1.3}
\begin{center}
\caption{Comparison between the best fit parameter values and
1$\sigma$ errors for the BBN constraints on
$\frac{\Delta\alpha}{\alpha_0}$, $\frac{\Delta <v>}{<v>_0}$ and
$\frac{\Delta m_e}{\left(m_e\right)_0}$, when one or two constants
are allowed to vary \cite{Mosquera07,Scoccola07}. We also present
the values of $\frac{\chi^2_{min}}{N-g}$ with $g=2$ for the case
where two constants are allowed to vary,  whereas $g=1$ when only
one fundamental constant is allowed to vary. We consider ${\rm
D}+^4{\rm He}$ data and $\eta_B$ fixed at the WMAP estimation.}
\label{resumenbbn}
\begin{tabular}{|c|c|c|c|c|c|}
\hline &\multicolumn{5}{c|}{Time variation of} \\ \cline{2-6}
Parameter&$\alpha$ and $<v>$  & $\alpha$ and $m_e$ & $\alpha$ &
$<v>$& $m_e$ \\ \hline

$\frac{\Delta\alpha}{\alpha_0}\pm \sigma$ &
$-0.030_{-0.030}^{+0.035}$&$-0.036_{-0.053}^{+0.052}$&$-0.020 \pm
0.007 $&---&---\\ \hline

$\frac{\Delta <v>}{<v>_0}\pm \sigma$ &
$-0.002_{-0.008}^{+0.007}$&---&---&$0.004 \pm 0.002$ &---\\ \hline

$\frac{\Delta m_e}{\left(m_e\right)_0}\pm \sigma $
&---&$0.020_{-0.064}^{+0.066}$&---&---&$-0.024 \pm 0.008$ \\
\hline

${\chi^2_{min}}/{\left(N-g\right)}$ &$1.03$&$1.00$&$0.90$&$0.97$&$0.95$\\
\hline
\end{tabular}
\end{center}
\end{table}

\section{Bounds from CMB}
\label{cmb}

The cosmological parameters can be estimated by an analysis of the
Cosmic Microwave Background (CMB) radiation, which gives information
about the physical conditions in the Universe just before decoupling
of matter and radiation.

\begin{figure}[!ht]
\begin{center}
\includegraphics[scale=0.50,angle=-90]{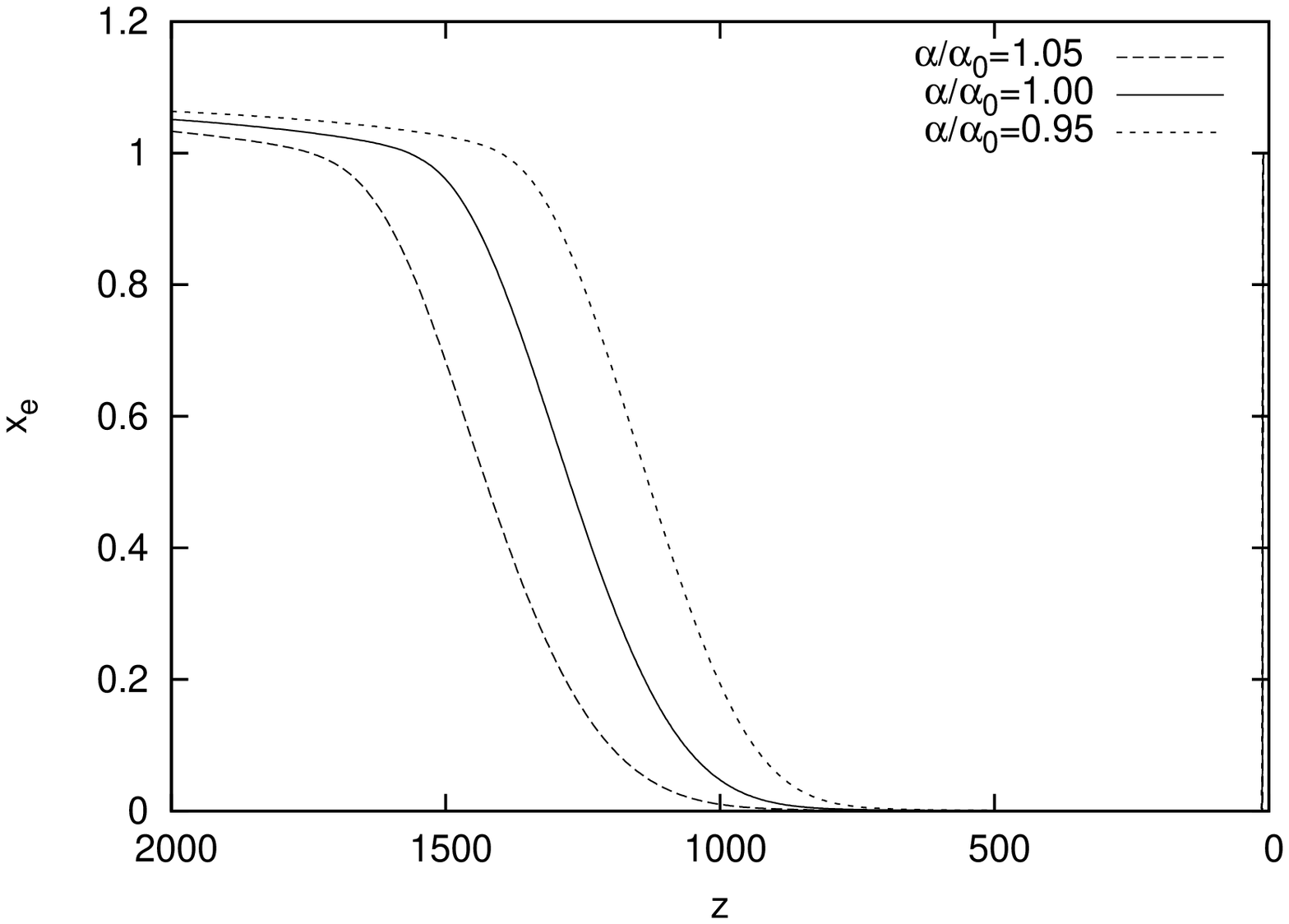}
\includegraphics[scale=0.50,angle=-90]{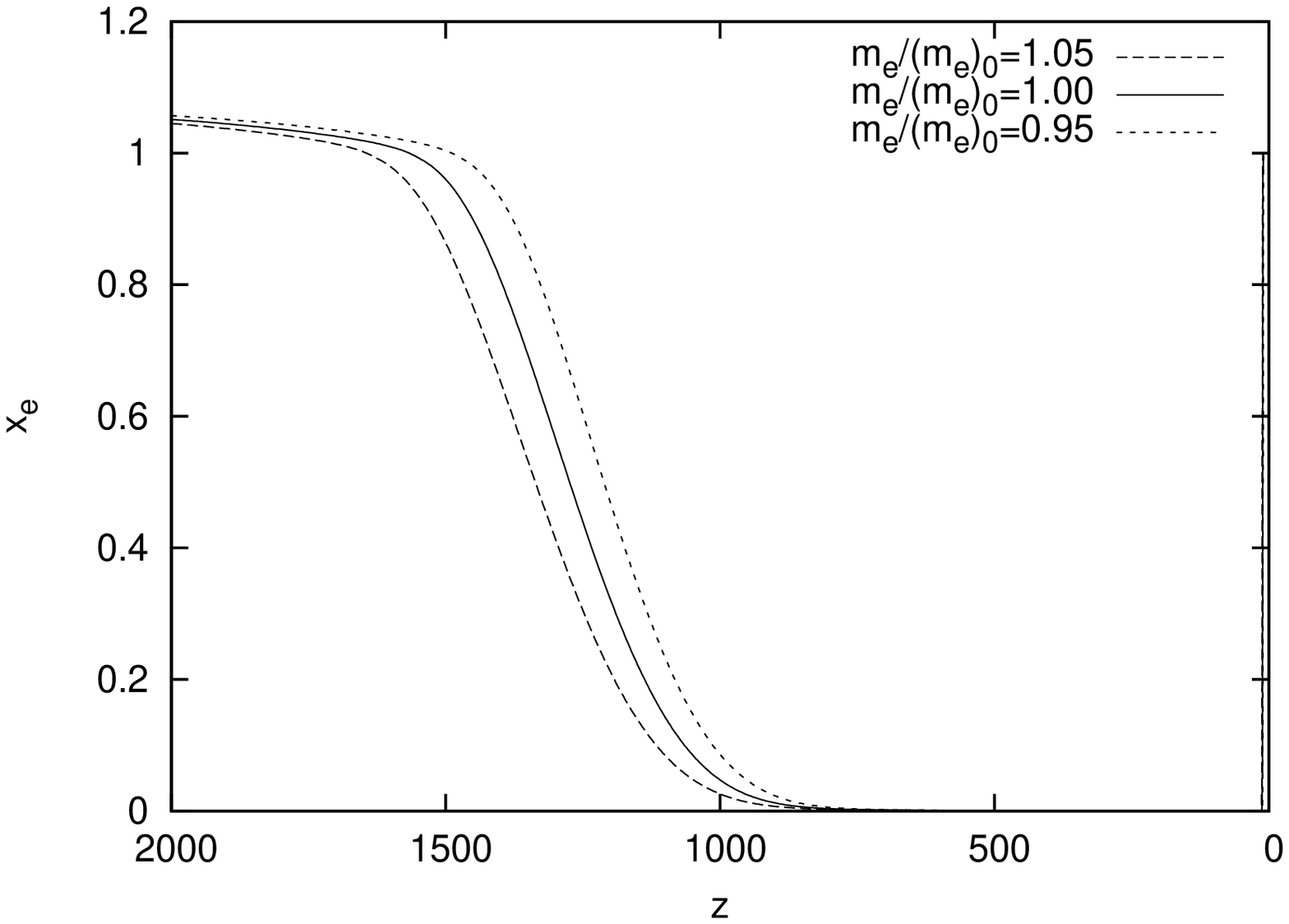}
\end{center}
\caption{Ionization history as a function of redshift, for different
  values of $\alpha$ (left panel) and $m_e$ (right panel) at
  recombination time.}
\label{ionization_history}
\end{figure}

The variation of fundamental constants affects the physics during
recombination (see appendix \ref{apendice_recombinacion} for
details). At this stage of the Universe history, the only
consequence of the time variation of $<v>$ is a variation in $m_e$.
The main effect of $\alpha$ and $m_e$ variations is the shift of the
epoch of recombination to higher $z$ as $\alpha$ or $m_e$ increases.
This is easy to understand since the binding energy $B_n$ scales as
$\alpha^2 m_e$, so photons should have higher energy to ionize
hydrogen atoms. In Figs.~{\ref{ionization_history}} we show how the
ionization history is affected by changes in $\alpha$ and in $m_e$,
in a flat universe with cosmological parameters $(\Omega_bh^2,
\Omega_{CDM}h^2, h, \tau) = (0.0223, 0.1047, 0.73, 0.09)$. When
$\alpha$ and/or $m_e$ have higher values than the present ones,
recombination occurs earlier (higher redshifts). The ionization
history is more sensitive to $\alpha$ than to $m_e$ because of the
$B_n$ dependence on this constants.

The most efficient thermalizing mechanism for the photon gas in the
early universe is Thomson scattering on free electrons. Therefore,
another important effect produced by the variation of fundamental
constants, is a shift in the Thomson scattering cross section
$\sigma_{T}$, which is proportional to $m_e^{-2}\alpha^2$.

\begin{figure}[!ht]
\begin{center}
\includegraphics[scale=0.50,angle=-90]{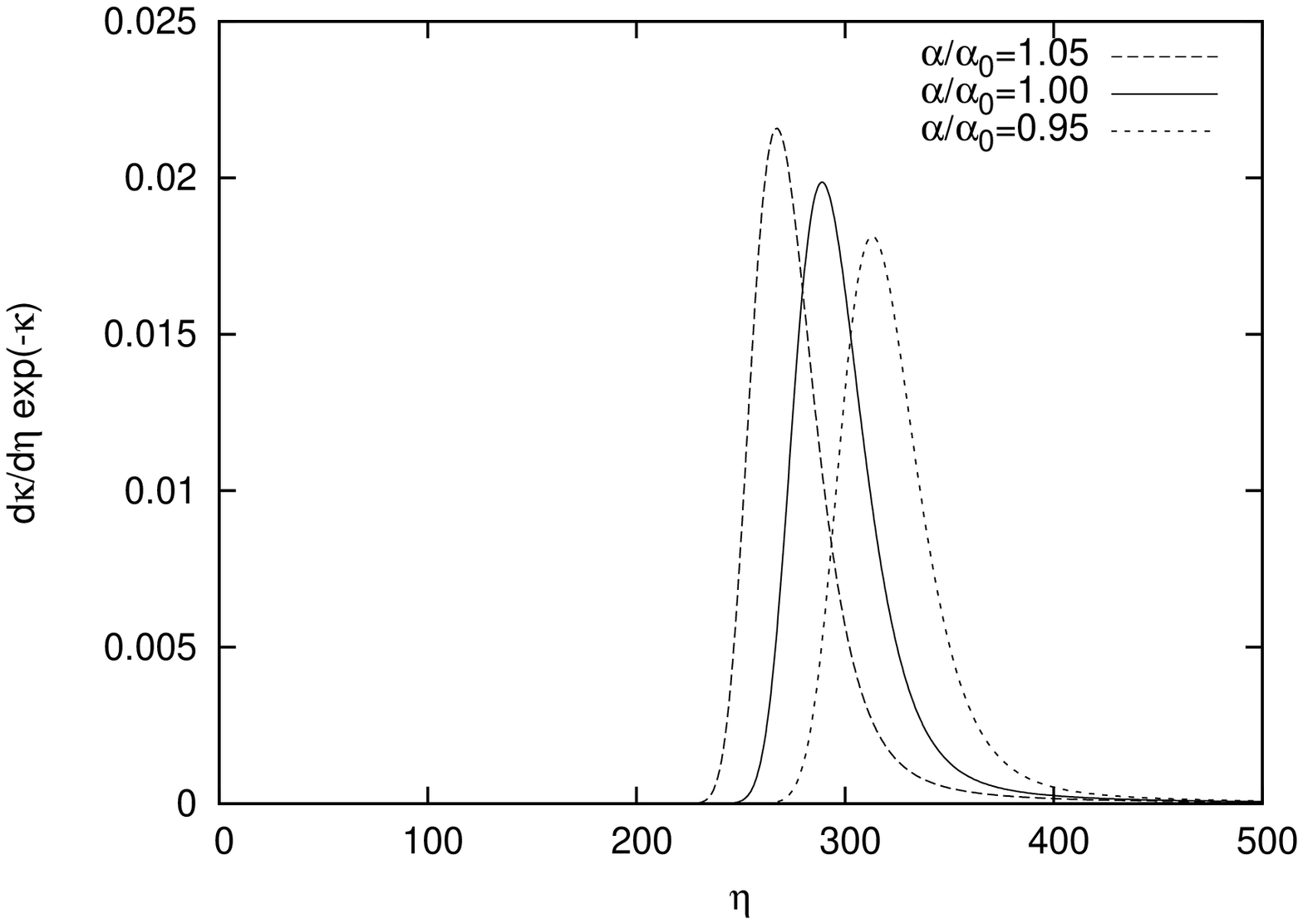}
\includegraphics[scale=0.50,angle=-90]{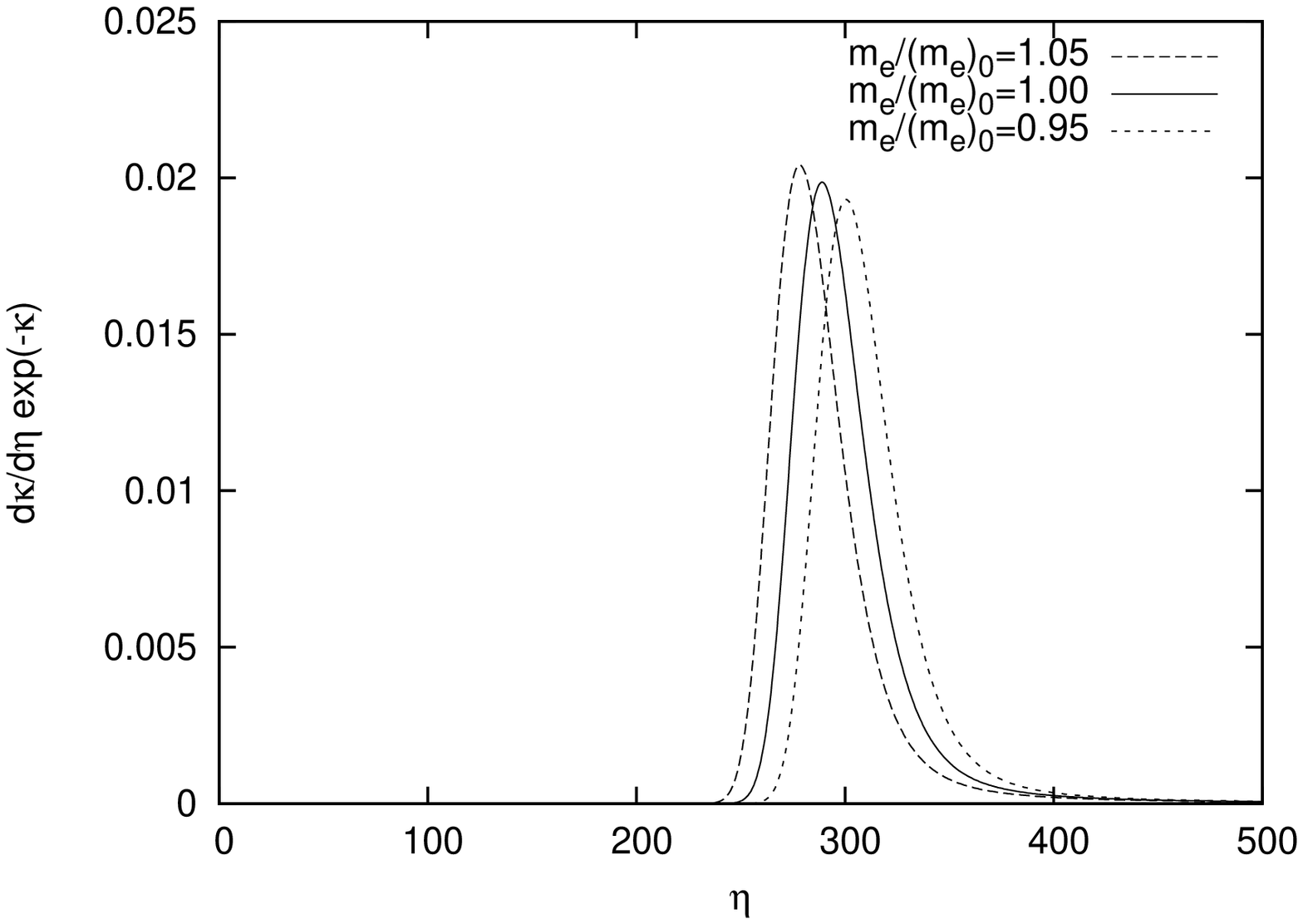}
\end{center}
\caption{Visibility function as a function of conformal time in Mpc, for different
  values of $\alpha$ (left panel) and $m_e$ (right panel).}
\label{visibility_function}
\end{figure}

The visibility function, which measures the differential probability
that a photon last scattered at conformal time $\eta$, depends on
$\alpha$ and $m_e$. This function is defined as
\begin{equation}
g(\eta)=e^{-\kappa}\frac{d \kappa}{d\eta} \, ,\qquad \mathrm{where}
\qquad \frac{d \kappa}{d\eta}=x_{e}n_{p} a  \sigma_{T} \,
\end{equation}
is the differential optical depth of photons due to Thomson
scattering, $n_{p}$ is the total number density of protons (both
free and bound), $x_{e}$ is the fraction of free electrons, and $a$
is the scale factor. The strongest effect of variations of $\alpha$
and $m_e$ on the visibility function occurs due to the alteration of
the ionization history $x_{e}(\eta)$. In
Fig.~\ref{visibility_function} we show that if $\alpha$ and/or $m_e$
were smaller (larger) at recombination than their present values,
the peak in the visibility function would shift towards smaller
(larger) redshifts, and its width would slightly increase
(decrease).

\begin{figure}[!ht]
\psfrag{angularl}{$\ell$}
\psfrag{powerCl}{$\ell (\ell +1) C_{\ell}/2\pi$}
\begin{center}
\includegraphics[scale=0.50,angle=-90]{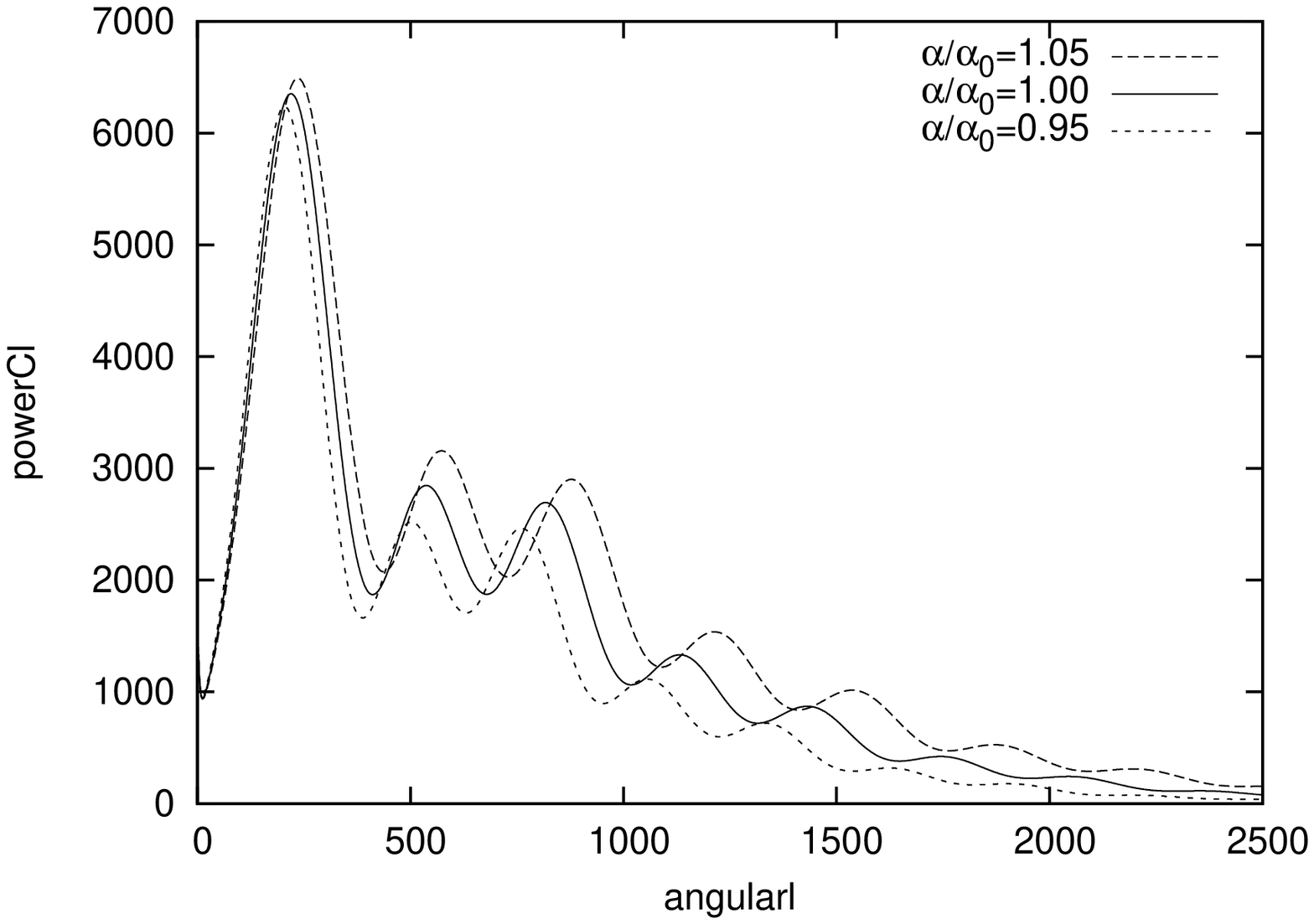}
\includegraphics[scale=0.50,angle=-90]{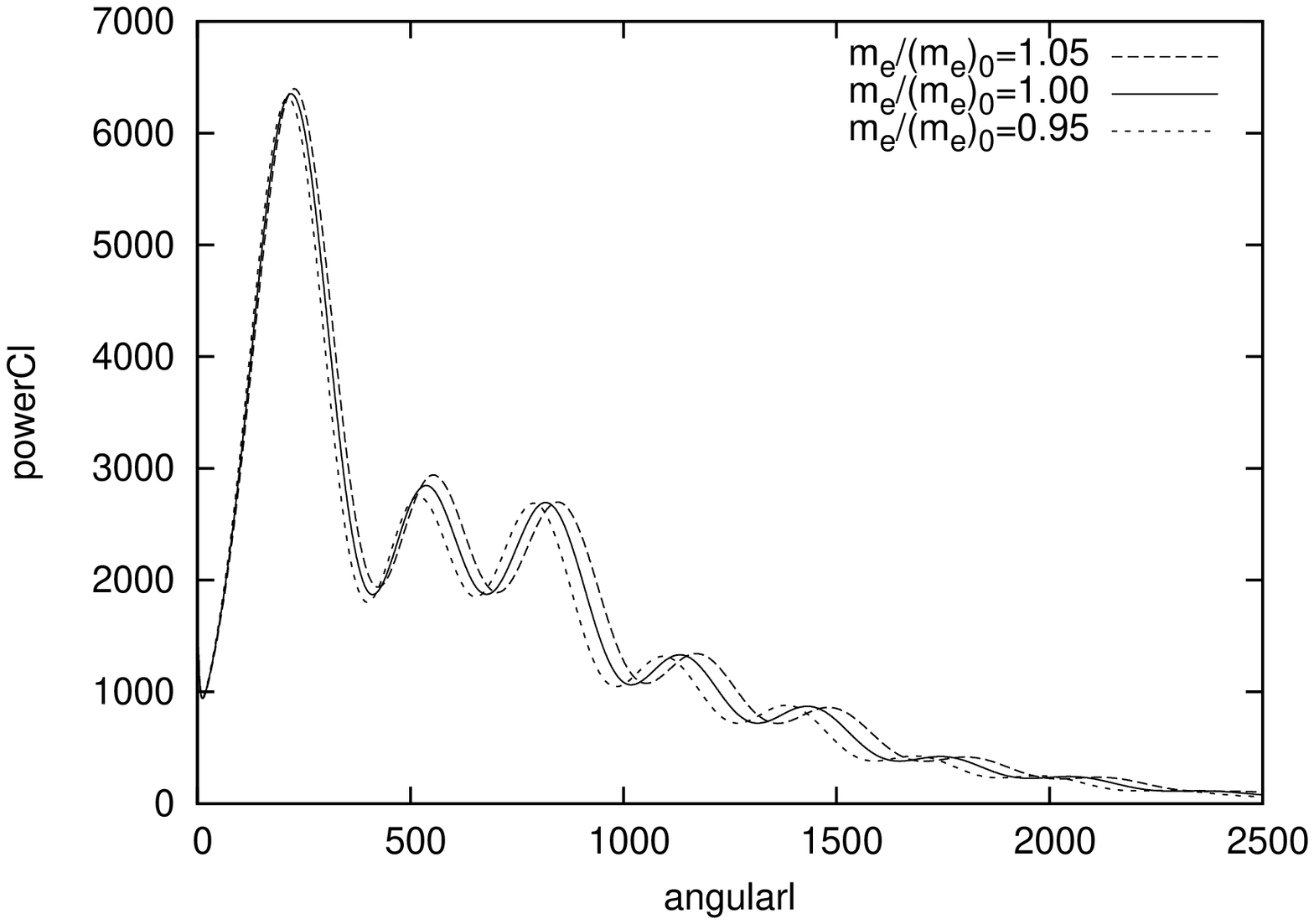}
\end{center}
\caption{The spectrum of CMB fluctuations for different
  values of $\alpha$ (left panel) and $m_e$ (right panel).}
\label{Cls_alfa_emasa}
\end{figure}

The signatures on the CMB angular power spectrum due to varying
fundamental constants are similar to those produced by changes in
the cosmological parameters, i.e. changes in the relative amplitudes
of the Doppler peaks and a shift in their positions. Indeed, an
increase in $\alpha$ or $m_e$ leads to a higher redshift of the
last-scattering surface, which corresponds to a smaller sound
horizon. The position of the first peak ($\ell_{1}$) is inversely
proportional to the latter, so a larger $\ell_{1}$ results.  Also a
larger early integrated Sach-Wolfe effect is produced, making the
first Doppler peak higher. Moreover, an increment in $\alpha$ or
$m_e$ decreases the high-$\ell$ diffusion damping, which is due to
the finite thickness of the last-scattering surface, and thus,
increases the power on very small scales \cite{BCW01,LHZ01,AV00}.
All these effects are illustrated in Figs.~\ref{Cls_alfa_emasa}.

To put constraints on the variation of $\alpha$ and $<v>$ during
recombination time, we performed a statistical analysis using data
from the WMAP 3-year temperature and temperature-polarization power
spectrum \cite{wmap3}, and other CMB experiments such as CBI
\cite{CBI04}, ACBAR \cite{ACBAR02}, and BOOMERANG
\cite{BOOM05_polar,BOOM05_temp}, and the power spectrum of the
2dFGRS \cite{2dF05}. We consider a spatially-flat cosmological model
with adiabatic density fluctuations, and the following parameters:
\begin{equation}
P=\left(\Omega_B h^2, \Omega_{CDM} h^2, \Theta, \tau, \frac{\Delta
\alpha}{\alpha_0}, \frac{\Delta <v>}{<v>_0}, n_s, A_s\right)\, \, ,
\end{equation}
where $\Omega_{CDM} h^2$ is the dark matter density in units of the
critical density, $\Theta$ gives the ratio of the comoving sound
horizon at decoupling to the angular diameter distance to the
surface of last scattering, $\tau$ is the reionization optical
depth, $n_s$ the scalar spectral index and $A_s$ is the amplitude of
the density fluctuations.

\begin{figure}[!ht]
\begin{center}
\includegraphics[scale=1.3,angle=0]{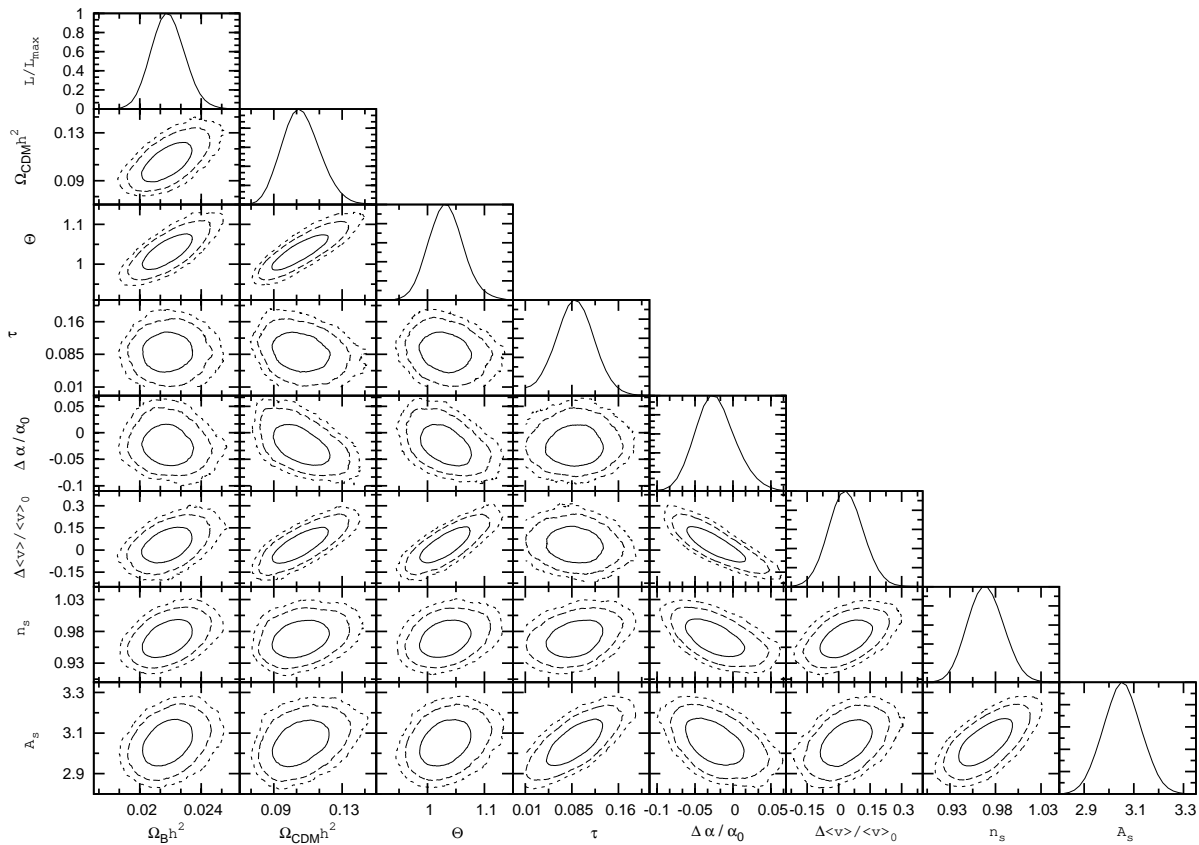}
\end{center}
\caption{Marginalized posterior distributions obtained with CMB
data, including the WMAP 3-year data release plus 2dFGRS power
spectrum. The diagonal shows the posterior distributions for
individual parameters, the other panels shows the 2D contours for
pairs of parameters, marginalizing over the others.}
\label{resulcmb}
\end{figure}

The parameter space was explored using the Markov Chain Monte Carlo
method implemented in the CosmoMC code of ref.  \cite{LB02} which
uses CAMB \cite{LCL00} to compute the CMB power spectra and RECFAST
\cite{recfast} to solve the recombination equations. We modified
these numerical codes in order to include the possible variation of
$\alpha$ and $<v>$ (or $m_e$) at recombination. We ran 8 Markov
chains and followed the convergence criterion of ref.
\cite{Raftery&Lewis} to stop them when $R-1<0.0149$. Results are
shown in Table \ref{tablacmb} and Figure \ref{resulcmb}.

\begin{table}[!ht]
\begin{center}
\renewcommand{\arraystretch}{1.3}
\caption{Mean values and 1$\sigma$ errors for the parameters
including $\alpha$ and $<v>$ variation. For comparison, results
where only one fundamental constant is allowed to vary are also
shown \cite{Mosquera07,Scoccola07}. $H_0$ is in units of ${\rm km \,
\, s^{-1} \, \, Mpc^{-1}}$.} \label{tablacmb}
\begin{tabular}{|c|c|c|c|}
\hline
Parameter & $\alpha$ and $<v>$ variation  & $\alpha$ variation & $<v>$ variation\\
\hline $\Omega_B h^2$ & $0.0218 \pm 0.0010$ & $0.0216 \pm 0.0009$ &
$0.0217 \pm 0.0010$  \\ \hline $\Omega_{CDM} h^2$ & $0.106 \pm
0.011$ & $0.102 \pm 0.006$ & $0.101 \pm 0.009$  \\ \hline $\Theta$ &
$1.033_{-0.029}^{+0.028}$  & $1.021 \pm 0.017$ & $1.020 \pm 0.025$
\\ \hline $\tau$ & $0.090 \pm 0.014$ & $0.092 \pm 0.014$  &
$0.091_{-0.014}^{+0.013}$ \\ \hline $\Delta \alpha / \alpha_0$ &
$-0.023 \pm 0.025$ & $-0.015 \pm 0.012$  & --- \\ \hline $\Delta
<v>/<v>_0$ & $0.036 \pm 0.078$   & --- & $-0.029 \pm 0.034$ \\
\hline $n_s$ &$0.970 \pm 0.019$ & $0.965 \pm 0.016$  & $0.960 \pm
0.015$ \\ \hline $A_s$ & $3.054 \pm 0.073$ &
$3.039_{-0.065}^{+0.064}$  & $3.020 \pm 0.064$ \\ \hline $H_0$ &
$70.4_{-6.8}^{+6.6}$ & $67.7_{-4.6}^{+4.7}$  & $68.1_{-6.0}^{+5.9}$
\\ \hline
\end{tabular}
\end{center}
\end{table}

It is noticeable the strong degeneracies that exist between
$\frac{\Delta <v>}{<v>_0}$ and $\Omega_{CDM} h^2$, $\frac{\Delta
<v>}{<v>_0}$ and $\Theta$, $\frac{\Delta \alpha}{\alpha_0}$ and
$n_s$, and also between $\frac{\Delta \alpha}{\alpha_0}$ and
$\frac{\Delta <v>}{<v>_0}$. The values obtained for $\Omega_B h^2$,
$h$, $\Omega_{CDM} h^2$, $\tau$, and $n_s$ agree, within 1$\sigma$,
with those of WMAP team \cite{wmap3}, where no variation of $\alpha$
nor $<v>$ is considered. It is interesting to note that our results
for the cosmological parameters are similar to those obtained
considering the variation of one constant at each time
\cite{Mosquera07,Scoccola07}. Our results are consistent within
1$\sigma$ with no variation of $\alpha$ and $<v>$ at recombination.

\begin{figure}[!ht]
\begin{center}
\includegraphics[scale=0.5,angle=-90]{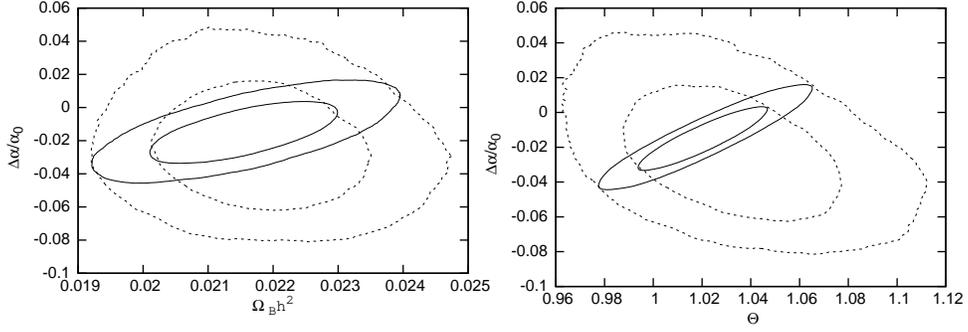}
\end{center}
\caption{1$\sigma$ and 2$\sigma$ contour levels. Dotted line:
variation of $\alpha$ and $<v>$; solid line: only $\alpha$
variation. The cosmological parameters are free to vary in both
cases.} \label{multi_alfa}
\end{figure}

\begin{figure}[!ht]
\begin{center}
\includegraphics[scale=0.5,angle=-90]{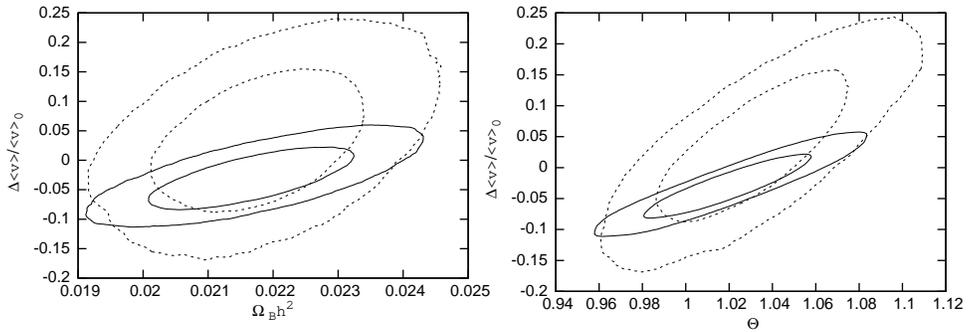}
\end{center}
\caption{1$\sigma$ and 2$\sigma$ contour levels. Dotted line:
variation of $\alpha$ and $<v>$; solid line: only $<v>$ variation.
The cosmological parameters are free to vary in both cases.}
\label{multi_emasa}
\end{figure}

In Figures \ref{multi_alfa} and \ref{multi_emasa} we compare the
degeneracies that exist between different cosmological parameters
and the fundamental constants when one or both constants are allowed
to vary. In any case, the allowable region in the parameter space is
larger when both fundamental constants are allowed to vary. This is
to be expected since when the parameter space has a higher dimension
the uncertainties in the parameters are larger. The correlations of
$\frac{\Delta \alpha}{\alpha_0}$ with the other cosmological
parameters change sign when $<v>$ is also allowed to vary. On the
contrary, the correlations of $\frac{\Delta <v>}{<v>_0}$ with
cosmological parameters do not change sign when $\alpha$ is also
allowed to vary.

When only one fundamental constant is allowed to vary, the
correlation between this constant and any particular cosmological
parameter has the same sign, no matter whether the fundamental
constant is $\alpha$ or $<v>$. This is because both constants enter
the same physical quantities. However, since the functional forms of
the dependence on $\alpha$ and $<v>$ are different, the best fit
mean values for the time variations of these fundamental constants
are different and the probability distribution is more extended in
one case than in the other. Nevertheless, in the cases when only one
constant is allowed to vary, it prefers a lower value than the
present one.

\section{Discussion}
\label{discusion}

In section \ref{nucleo}, we obtained bounds on the variation of
$\alpha$ and $<v>$ using the observational abundances of ${\rm D}$,
$^4{\rm He}$ and $^7{\rm Li}$. We performed different analyses: i)
we allow $\eta_B$ to vary and ii) we keep $\eta_B$ fixed. We also
performed the same analyses for two different estimations of the
dependence of the deuterium binding energy on the pion mass or the
Higgs vacuum expectation value: i) the obtained by Yoo and Scherrer
who considered the coefficient for the linear dependence obtained by
Bean and Savage \cite{BS03b}; ii) the obtained using the Reid
potential for the description of the nucleon-nucleon interaction and
without using chiral perturbation theory. The best fits for these
two different cases are consistent within 1$\sigma$. We found
reasonable fit for the variation of $\alpha$, $<v>$ and $\eta_B$ for
the whole data set and for the variation of $\alpha$, $<v>$, keeping
$\eta_B$ fixed, for the whole data set and also when we exclude one
group of data. We only found variation of the fundamental constants
when the $^7{\rm Li}$ abundance is included in the statistical
analysis.  We also calculated the light abundances, keeping $\eta_B$
fixed at the WMAP estimation, for different values of the dependence
of $\epsilon_D$ on the Higgs vacuum expectation value, inside the
range proposed in ref. \cite{EMG03}, and performed the statistical
analysis. These results are consistent within 1$\sigma$ with the
ones presented in section \ref{nucleo}.

We also considered the joint variation of $\alpha$ and $m_e$ with
$\eta_B$ variable and fixed at the WMAP estimation. In this case, we
also obtained reasonable fits for the whole data set. When the $^7{\rm
  Li}$ abundance was included in the fit, we obtained results
consistent with variation of fundamental constants (and $\eta_B$
consistent with the WMAP value). From a phenomenological point of
view, to vary $\alpha$ and $m_e$ solves the discrepancy between the
$^7{\rm Li}$ data, the other abundances and the WMAP estimate.
However, it is important to mention that the theoretical motivations
for $m_e$ being the varying fundamental constant are weak.

We have discussed in section \ref{nucleo} that there is still no
agreement within the astronomical community in the value of the
$^7{\rm Li}$ abundance. We think that more observations of $^7{\rm
Li}$ are needed in order to arrive to stronger conclusions. However,
if the present values of $^7{\rm Li}$ abundances are correct, we may
have insight into new physics and varying fundamental constants
would be a good candidate for solving the discrepancy between the
light elements abundances and the WMAP estimates.

In section \ref{cmb}, we calculated the time variation of $\alpha$
and $<v>$ (or $m_e$) with data from CMB observations and the final
2dFGRS power spectrum. In this analysis, we also allowed other
cosmological parameters to vary. We found no variation of $\alpha$
and $<v>$ within 1$\sigma$, and the values for the cosmological
parameters agree with those obtained by ref. \cite{wmap3} within
1$\sigma$.

\begin{figure}[!ht]
\begin{center}
\includegraphics[scale=0.35,angle=-90]{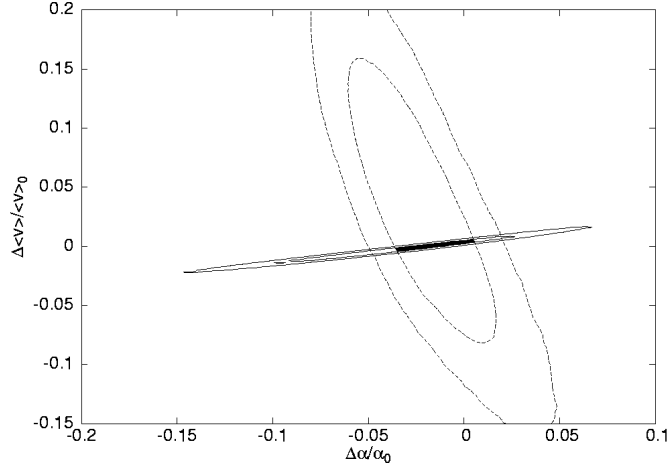}
\end{center}
\caption{2D contour levels for variation of $\alpha$ and $<v>$ from
BBN (solid line) and CMB (dotted line) data. } \label{final-v}
\end{figure}

In Figure \ref{final-v} we show the 2D contours for $\frac{\Delta
\alpha}{\alpha_0}$ and $\frac{\Delta <v>}{<v>_0}$ obtained from BBN
and CMB data. The correlation coefficients are $-0.82$ for CMB and
$0.77$ for BBN. There is a small region where the two contours
superpose which is consistent with null variation of both constants.
However, the results do not exclude the possibility that the
fundamental constants have values different from their present ones
but constant in the early universe. It is possible to obtain a
linear relationship (between $\frac{\Delta \alpha}{\alpha_0}$ and
$\frac{\Delta <v>}{<v>_0}$) from the BBN and CMB contours:
\begin{eqnarray}
\frac{\Delta <v>}{<v>_0}&=& a_{BBN} \frac{\Delta
\alpha}{\alpha_0}+b_{BBN} \hskip 1cm {\rm for \, \, BBN,} \\
\frac{\Delta <v>}{<v>_0} &=& a_{CMB} \frac{\Delta \alpha}{\alpha_0}
+b_{CMB} \hskip 1cm {\rm for \, \, CMB,}
\end{eqnarray}
where $a_{BBN}= 0.181\pm 0.003$, $b_{BBN}=0.0046\pm 0.0002$,
$a_{CMB}=-3.7^{+0.1}_{-0.5}$ and $b_{CMB}=-0.053^{+0.009}_{-0.027}$.

\begin{figure}[!ht]
\begin{center}
\includegraphics[scale=0.35,angle=-90]{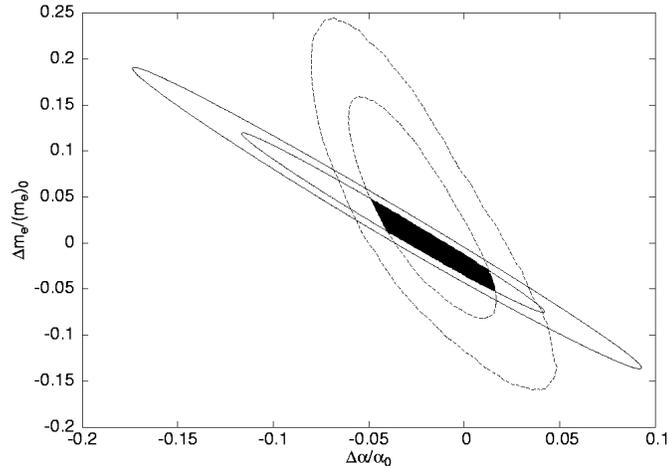}
\end{center}
\caption{2D contour levels for variation of $\alpha$ and $m_e$ from
BBN (solid line) and CMB (dotted line) data. } \label{final-me}
\end{figure}

Figure \ref{final-me} show the 2D contours for $\frac{\Delta
\alpha}{\alpha_0}$ and $\frac{\Delta m_e}{\left(m_e\right)_0}$
obtained from BBN and CMB data. In this case, the correlation
coefficients are $-0.82$ for CMB and $-0.91$ for BBN. A
phenomenological relationship between the variation of the
fundamental constants $\alpha$ and $m_e$ can be obtained by
adjusting a linear function. These two linear fits are different for
both cases:
\begin{eqnarray}
\frac{\Delta m_e}{\left(m_e\right)_0}&=&c_{BBN}\frac{\Delta
\alpha}{\alpha_0}+d_{BBN}\hskip 1cm {\rm for \, \, BBN,} \\
\frac{\Delta m_e}{\left(m_e\right)_0} &=& c_{CMB} \frac{\Delta
\alpha}{\alpha_0} +d_{CMB} \hskip 1cm {\rm for \, \, CMB,}
\end{eqnarray}
where $c_{BBN}=-1.229\pm 0.008$, $d_{BBN}=-0.0234\pm 0.005$,
$c_{CMB}=-3.7^{+0.1}_{-0.5}$ and $d_{CMB}=-0.053^{+0.009}_{-0.027}$
(the time variation of $<v>$ during CMB has the same effects than
the variation of the electron mass).

It is important to point out that BBN degeneration suggests
phenomenological relationships between the variations of both
constants, while the CMB contours are not thin enough to assure any
conclusion.

Our results suggest that the model used by Ichikawa et al.
\cite{ichi06} where the variation of fundamental constants is driven
by the time evolution of a dilaton field can be discarded, since
these models predict $m_e \simeq \alpha^{1/2}$.

\section{Summary and Conclusion}
\label{conclusion}

In this work we have studied the joint time variation of the fine
structure constant and the Higgs expectation value and the joint
variation of $\alpha$ and $m_e$ in the early Universe. We used the
observational abundances of ${\rm D}$, $^4{\rm He}$ and $^7{\rm Li}$
to put bounds on the joint variation of $\alpha$ and $<v>$ and on
the joint variation of $\alpha$ and $m_e$ during primordial
nucleosynthesis. We used the three year WMAP data together with
other CMB experiments and the 2dfGRS power spectrum to put bounds on
the variation of $\alpha$ and $<v>$ (or $m_e$) at the time of
neutral hydrogen formation.

From our analysis we arrive to the following conclusions:
\begin{enumerate}
\item{The consideration of different values of $\frac{\partial
\epsilon_D}{\partial m_\pi}$ leads to similar constraints on the time
variation of the fundamental constants.}

\item{We obtain non null results for the joint variation of $\alpha $
  and $<v>$  at $ 6 \sigma$ in two cases: i) when $\eta_B$ is allowed
  to vary and all abundances are included in the data set used to
  perform the fit and ii) when only $\He$ and $\Li$ are included in
  the data set used to perform the fit and $\eta_B$ is fixed to the
  WMAP estimation. In the first case, the obtained value of $\eta_B$
  is inconsistent with the WMAP estimation.}

\item{We obtain non null results for the joint variation of $\alpha$
  and $m_e$ at $ 6 \sigma$ in two cases: i) when $\eta_B$ is allowed
  to vary and all abundances are included in the data set used to
  perform the fit and ii) when only $\He$ and $\Li$ are included in
  the data set used to perform the fit and $\eta_B$ is fixed to the
  WMAP estimation. In the first case, the obtained value of $\eta_B$
  is inconsistent with the WMAP estimation.}

{\item We also obtain non null results for the joint variation of
  $\alpha$ and $<v>$ and $\alpha$ and $m_e$ when all abundances are
  included in the data set and $\eta_B$ is fixed to the WMAP
  estimation. The statistical significance of these results is too low
  to claim a variation of fundamental constants.}

\item{Excluding $^7{\rm Li}$ abundance from the data set
used to perform the fit, and keeping $\eta_B$ fixed, we find results
that are consistent with no variation of fundamental constants within
1$\sigma$.}

\item{The bounds obtained using data from CMB and 2dFGRS are
consistent with null variation of $\alpha$ and $<v>$ (or $m_e$) at
recombination within 1$\sigma$.}

\item{We find phenomenological relationships for the
variations of $\alpha$ and $<v>$, and for the variations of $\alpha$
and $m_e$, at the time of primordial  nucleosynthesis and at the time
of recombination. All the phenomenological relationships correspond
to linear fits.}

\item{From our phenomenological approach, it follows that the relationship
between the variations of the two pairs of constants considered in
this paper is different at the time of nucleosynthesis that at the
time of neutral hydrogen formation.}

\item{The dilaton model proposed by Ichikawa et al. \cite{ichi06} can be
discarded.}
\end{enumerate}

\section*{{\bf Acknowledgments}}

Support for this work was provided by Project G11/G071, UNLP and PIP
5284 CONICET. The authors would like to thank Andrea Barral,
Federico Bareilles, Alberto Camyayi and Juan Veliz for technical and
computational support. The authors would also like to thank Ariel
Sanchez for support with CosmoMC. MEM wants to thank O. Civitarese
and S. Iguri for the interesting and helpful discussions. CGS gives
special thanks to Licia Verde and Nelson Padilla for useful
discussion.

\appendix

\section{Physics at BBN}
\label{correcciones}

We discuss the dependencies on $\alpha$, $<v>$, and $m_e$ of the
physical quantities involved in the calculation of the abundances of
the light elements. We also argue how these quantities are modified in
the Kawano code.

\subsection{Variation of the fine structure constant}

The variation of the fine structure constant affects several physical
quantities relevant during BBN. These quantities are the cross
sections, the Q-values of reaction rates, the light nuclei masses,
and the neutron-proton mass difference (along with the neutrons and
protons initial abundances and the $n\leftrightarrow p$ reaction
rates).

The cross sections were modified following refs.
\cite{ichi06,Nollet,LMV06,Iguri99} and replacing $\alpha$ by
$\alpha_0 \left(1+\frac{\Delta \alpha}{\alpha_0}\right)$ in the
numerical code. The Q-values of reaction rates where modified
following ref. \cite{LMV06}.

To consider the effect of the variation of the fine structure
constant upon the light nuclei masses, we adopted:
\begin{eqnarray}
\frac{\Delta m_x}{\left(m_x\right)_0}&=& P
\frac{\Delta\alpha}{\alpha_0}\,\, ,
\end{eqnarray}
where $P$ is a constant of the order of $10^{-4}$ (see ref.
\cite{LMV06} for details) and $m_x$ is the mass of the nuclei $x$.
These changes affect all of the reaction rates, their Q-values and
their inverse coefficients.

If the fine structure constant varies with time, the neutron-proton
mass difference also changes. Following ref. \cite{Epele91b}:
\begin{eqnarray}
\frac{\delta \Delta m_{np}}{\Delta m_{np}} &=&-0.587 \frac{\Delta
\alpha}{\alpha_0} \, \, .
\end{eqnarray}
This modifies the $n\leftrightarrow p$ and the neutrons and protons
initial abundances. The $n\to p$ reaction rate is calculated
by:
\begin{eqnarray}
\label{lambdanp} \lambda_{n\to p}= K \int_{m_e}^{\infty}
{\rm d}E_e \frac{E_e p_e}{1+e^{E_e/T_\gamma}} \frac{\left(E_e +
\Delta m_{np}\right)^2 }{1+e^{-\left(E_e+ \Delta m_{np}\right)
/T_\nu -\xi_l}} + K \int_{m_e}^{\infty} {\rm d}E_e \frac{E_e
p_e}{1+e^{-E_e/T_\gamma}}\frac{\left(E_e - \Delta
m_{np}\right)^2}{1+e^{\left(E_e-\Delta m_{np}\right) /T_\nu
+\xi_l}}\, \, ,
\end{eqnarray}
where $K$ is a normalization constant proportional to $G_F^2$, $E_e$
and $p_e$ are the electron energy and momentum respectively,
$T_\gamma$ and $T_\nu$ are the photon and neutrino temperature and
$\xi_l$ is the ratio between the neutrino chemical potential and the
neutrino temperature. In order to include the variation of $\Delta
m_{mp}$ we replace this quantity by $\Delta m_{mp}\left(1+
\frac{\delta \Delta m_{np}}{\Delta m_{np}}\right)$ in Kawano code.
The neutrons and protons initial abundances are calculated by:
\begin{eqnarray}
\label{ynyp}
Y_n&=& \frac{1}{1+e^{\Delta m_{np}/T +\xi}}\, \, ,\\
Y_p&=&  \frac{1}{1+e^{-\Delta m_{np}/T -\xi}}\, \, .
\end{eqnarray}

\subsection{Variation of the electron mass}

If the electron mass can have a different value than the present one
during primordial nucleosynthesis, the sum of the electron and
positron energy densities, the sum of the electron and positron
pressures and the difference of the electron and positron number
densities must be modified in order to include this change. These
quantities are calculated in Kawano code as:
\begin{eqnarray}
\label{rhoe} \rho_{e^-}+ \rho_{e^+}&=& \frac{2}{\pi^2}
\frac{\left(m_e c^2\right)^4}{\left(\hbar c \right)^3} \sum_n
(-1)^{n+1} {\rm cosh} \left(n\phi_e\right) M(nz)\, \, ,\\
\label{pe} \frac{p_{e^-}+ p_{e^+}}{c^2}&=& \frac{2}{\pi^2}
\frac{\left(m_e c^2\right)^4}{\left(\hbar c \right)^3} \sum_n
\frac{(-1)^{n+1}}{nz} {\rm cosh} \left(n\phi_e\right) N(nz)\, \, ,\\
\label{ne} \frac{\pi^2}{2}\left[\frac{\hbar c^3}{m_e c^2}\right]^3
z^3\left(n_{e^-}-n_{e^+}\right)&=& z^3 \sum_n (-1)^{n+1}{\rm
sinh}\left(n\phi_e\right) L(nz)\, \, ,
\end{eqnarray}
where $z=\frac{m_e c 2}{k T_\gamma}$, $\phi_e$ is the electron
chemical potential and $L(z)$, $M(z)$ and $N(z)$ are combinations of
the modified Bessel function $K_i(z)$ \citep{Kawano88,Kawano92}. The
change in these quantities affects their derivatives and the
expansion rate through the Friedmann equation:
\begin{eqnarray}
H^2&=&\frac{8 \pi}{3} G\left(\rho_T +\frac{\Lambda}{3}\right)\, \, ,
\end{eqnarray}
where $G$ is the Newton constant, $\Lambda$ is the cosmological
constant and
\begin{eqnarray}
\rho_T &=&\rho_\gamma+\rho_{e^-}+ \rho_{e^+}+\rho_\nu+\rho_b \, \, ,
\end{eqnarray}
The $n\leftrightarrow p$ reaction rates (see Eq.(\ref{lambdanp}))
and the weak decay rates of heavy nuclei are also modified if the
electron mass varies with time.

It is worth while mentioning that the most important changes in the
primordial abundances (due to a change in $m_e$) arrive from the
change in the weak rates rather than from the change in the
expansion rate \citep{YS03}.

\subsection{Variation of the Higgs vacuum expectation value}

If the value of $<v>$ during BBN is different than the present
value, the electron mass, the Fermi constant, the neutron-proton
mass difference and the deuterium binding energy take different
values than the current ones. The electron mass is proportional to
the Higgs vacuum expectation value, then
\begin{eqnarray}
\frac{\Delta m_e}{\left(m_e\right)_0}&=&\frac{\Delta <v>}{<v>_0}\,
\, .
\end{eqnarray}
The Fermi constant is proportional to $<v>^{-2}$ \citep{dixit88};
this dependence affects the $n\leftrightarrow p$ reaction rates. The
neutron-proton mass difference changes by \citep{Epele91b}
\begin{eqnarray}
\frac{\delta \Delta m_{np}}{\Delta m_{np}} &=&1.587 \frac{\Delta
<v>}{<v>_0}\, \, ,
\end{eqnarray}
affecting $n\leftrightarrow p$ reaction rates (see
Eq.(\ref{lambdanp})) and the initial neutron and proton abundances
(see Eq.(\ref{ynyp})).

The deuterium binding energy must be corrected by
\begin{eqnarray}
\frac{\Delta \epsilon_D}{\left( \epsilon_D\right)_0} &=&\kappa
\frac{\Delta <v>}{<v>_0}\, \, , 
\end{eqnarray}
where $\kappa$ is a model dependent constant. This constant can be
found: i) using chiral perturbation theory, as was done by Beane
and Savage \cite{BS03}; ii) using effective potentials to
describe the nucleon-nucleon interaction. This correction affects
the initial value of the deuterium abundance
\begin{eqnarray}
Y_d&=&\frac{Y_n Y_p e^{11.605 \epsilon_D/T_9}}{0.471×
10^{-10}T_9^{3/2}}\, \, ,
\end{eqnarray}
where $T_9$ is the temperature in units of $10^9 {\rm K}$, and
$\epsilon_D$ is in MeV.

\section{Physics at recombination}
\label{apendice_recombinacion}

During recombination epoch, the ionization fraction, $x_e= n_e/n$
(where $n_e$ and $n$ are the number density of free electrons and of
neutral hydrogen, respectively), is determined by the balance between
photoionization and recombination.

In this paper, we solved the recombination equations using RECFAST
\cite{recfast}, taking into account all of the dependencies on
$\alpha$ and $m_e$ \cite{Martins02,Rocha03,ichi06,YS03}. To get a
feeling of the dependencies of the physical quantities relevant during
recombination, we consider here the Peebles recombination scenario
\cite{Peebles68}.  The recombination equation is
\begin{equation}
-\frac{d}{dt}\left(\frac{n_e}{n}\right) = C \left( \frac{\alpha_c
n_e^2}{n} -\beta_c \frac{n_{1s}}{n} e^{-(B_1 - B_2)/kT} \right)  \, \, ,
\end{equation}
where
\begin{equation}
C= \frac{\left(1 + K \Lambda_{2s,1s} n_{1s}\right)}{\left(1 + K
(\beta_c + \Lambda_{2s,1s}) n_{1s}\right)}
\end{equation}
is the Peebles factor, which inhibits the recombination rate due to
the presence of Lyman-$\alpha$ photons, $n_{1s}$ is the number density
of hydrogen atoms in the ground state, and $B_{n}$ is the binding
energy of hydrogen in the $n$th principal quantum number.  The
redshift of the Lyman-$\alpha$ photons is $K = \frac{\lambda_\alpha^3
\ a}{8 \pi \dot a}$, with $\lambda_\alpha = \frac{8 \pi \hbar c}{3
B_1}$, and $\Lambda_{2s,1s}$ is the rate of decay of the $2s$ excited
state to the ground state via 2-photon emission, and scales as
$\alpha^8 m_e$. Recombination directly to the ground state is strongly
inhibited, so the \emph{case B} recombination takes place. The
\emph{case B} recombination coefficient $\alpha_c$ is proportional to
$\alpha^3 m_e^{-3/2}$. The photoionization coefficient depends on
$\alpha_c$, but it also has an additional dependence on $m_e$,
\begin{equation}
\beta_c = \alpha_c \left(\frac{2 \pi {m_e} k T}{h^2}\right)^{3/2}
e^{-B_2 / kT} \, \, .\nonumber
\end{equation}
The most important effects of changes in $\alpha$ and $m_e$ during
recombination are due to their influence upon Thomson scattering cross
section $\sigma _T = \frac{8\pi \ \hbar ^2}{3\ m_e^2 c^2}\alpha^2$,
and the binding energy of hydrogen $B_1={1\over2}\ \alpha^2 m_e c^2$.

\bibliography{bibliografia3}

\begin{thebibliography}{105}
\expandafter\ifx\csname natexlab\endcsname\relax\def\natexlab#1{#1}\fi
\expandafter\ifx\csname bibnamefont\endcsname\relax
  \def\bibnamefont#1{#1}\fi
\expandafter\ifx\csname bibfnamefont\endcsname\relax
  \def\bibfnamefont#1{#1}\fi
\expandafter\ifx\csname citenamefont\endcsname\relax
  \def\citenamefont#1{#1}\fi
\expandafter\ifx\csname url\endcsname\relax
  \def\url#1{\texttt{#1}}\fi
\expandafter\ifx\csname urlprefix\endcsname\relax\def\urlprefix{URL }\fi
\providecommand{\bibinfo}[2]{#2}
\providecommand{\eprint}[2][]{\url{#2}}

\bibitem[{\citenamefont{{Wu} and {Wang}}(1986)}]{Wu86}
\bibinfo{author}{\bibfnamefont{Y.}~\bibnamefont{{Wu}}} \bibnamefont{and}
  \bibinfo{author}{\bibfnamefont{Z.}~\bibnamefont{{Wang}}},
  \bibinfo{journal}{\prl} \textbf{\bibinfo{volume}{57}}, \bibinfo{pages}{1978}
  (\bibinfo{year}{1986}).

\bibitem[{\citenamefont{{Maeda}}(1988)}]{Maeda88}
\bibinfo{author}{\bibfnamefont{K.}~\bibnamefont{{Maeda}}},
  \bibinfo{journal}{Modern Physics. Letters A} \textbf{\bibinfo{volume}{31}},
  \bibinfo{pages}{243} (\bibinfo{year}{1988}).

\bibitem[{\citenamefont{{Barr} and {Mohapatra}}(1988)}]{Barr88}
\bibinfo{author}{\bibfnamefont{S.~M.} \bibnamefont{{Barr}}} \bibnamefont{and}
  \bibinfo{author}{\bibfnamefont{P.~K.} \bibnamefont{{Mohapatra}}},
  \bibinfo{journal}{\prd} \textbf{\bibinfo{volume}{38}}, \bibinfo{pages}{3011}
  (\bibinfo{year}{1988}).

\bibitem[{\citenamefont{{Damour} and {Polyakov}}(1994)}]{DP94}
\bibinfo{author}{\bibfnamefont{T.}~\bibnamefont{{Damour}}} \bibnamefont{and}
  \bibinfo{author}{\bibfnamefont{A.~M.} \bibnamefont{{Polyakov}}},
  \bibinfo{journal}{Nuclear Physics B} \textbf{\bibinfo{volume}{95}},
  \bibinfo{pages}{10347} (\bibinfo{year}{1994}).

\bibitem[{\citenamefont{{Damour}
  et~al.}(2002{\natexlab{a}})\citenamefont{{Damour}, {Piazza}, and
  {Veneziano}}}]{DPV2002a}
\bibinfo{author}{\bibfnamefont{T.}~\bibnamefont{{Damour}}},
  \bibinfo{author}{\bibfnamefont{F.}~\bibnamefont{{Piazza}}}, \bibnamefont{and}
  \bibinfo{author}{\bibfnamefont{G.}~\bibnamefont{{Veneziano}}},
  \bibinfo{journal}{\prl} \textbf{\bibinfo{volume}{89}},
  \bibinfo{pages}{081601} (\bibinfo{year}{2002}{\natexlab{a}}).

\bibitem[{\citenamefont{{Damour}
  et~al.}(2002{\natexlab{b}})\citenamefont{{Damour}, {Piazza}, and
  {Veneziano}}}]{DPV2002b}
\bibinfo{author}{\bibfnamefont{T.}~\bibnamefont{{Damour}}},
  \bibinfo{author}{\bibfnamefont{F.}~\bibnamefont{{Piazza}}}, \bibnamefont{and}
  \bibinfo{author}{\bibfnamefont{G.}~\bibnamefont{{Veneziano}}},
  \bibinfo{journal}{\prd} \textbf{\bibinfo{volume}{66}},
  \bibinfo{pages}{046007} (\bibinfo{year}{2002}{\natexlab{b}}).

\bibitem[{\citenamefont{{Youm}}(2001{\natexlab{a}})}]{Youm2001a}
\bibinfo{author}{\bibfnamefont{D.}~\bibnamefont{{Youm}}},
  \bibinfo{journal}{\prd} \textbf{\bibinfo{volume}{63}},
  \bibinfo{pages}{125011} (\bibinfo{year}{2001}{\natexlab{a}}).

\bibitem[{\citenamefont{{Youm}}(2001{\natexlab{b}})}]{Youm2001b}
\bibinfo{author}{\bibfnamefont{D.}~\bibnamefont{{Youm}}},
  \bibinfo{journal}{\prd} \textbf{\bibinfo{volume}{64}},
  \bibinfo{pages}{085011} (\bibinfo{year}{2001}{\natexlab{b}}).

\bibitem[{\citenamefont{{Palma} et~al.}(2003)\citenamefont{{Palma}, {Brax},
  {Davis}, and {van de Bruck}}}]{branes03a}
\bibinfo{author}{\bibfnamefont{G.~A.} \bibnamefont{{Palma}}},
  \bibinfo{author}{\bibfnamefont{P.}~\bibnamefont{{Brax}}},
  \bibinfo{author}{\bibfnamefont{A.~C.} \bibnamefont{{Davis}}},
  \bibnamefont{and} \bibinfo{author}{\bibfnamefont{C.}~\bibnamefont{{van de
  Bruck}}}, \bibinfo{journal}{\prd} \textbf{\bibinfo{volume}{68}},
  \bibinfo{pages}{123519} (\bibinfo{year}{2003}).

\bibitem[{\citenamefont{{Brax} et~al.}(2003)\citenamefont{{Brax}, {van de
  Bruck}, {Davis}, and {Rhodes}}}]{branes03b}
\bibinfo{author}{\bibfnamefont{P.}~\bibnamefont{{Brax}}},
  \bibinfo{author}{\bibfnamefont{C.}~\bibnamefont{{van de Bruck}}},
  \bibinfo{author}{\bibfnamefont{A.-C.} \bibnamefont{{Davis}}},
  \bibnamefont{and} \bibinfo{author}{\bibfnamefont{C.~S.}
  \bibnamefont{{Rhodes}}}, \bibinfo{journal}{Astrophysics and Space Science}
  \textbf{\bibinfo{volume}{283}}, \bibinfo{pages}{627} (\bibinfo{year}{2003}).

\bibitem[{\citenamefont{Kaluza}(1921)}]{Kaluza}
\bibinfo{author}{\bibfnamefont{T.}~\bibnamefont{Kaluza}},
  \bibinfo{journal}{Sitzungber. Preuss. Akad. Wiss.K}
  \textbf{\bibinfo{volume}{1}}, \bibinfo{pages}{966} (\bibinfo{year}{1921}).

\bibitem[{\citenamefont{Klein}(1926)}]{Klein}
\bibinfo{author}{\bibfnamefont{O.}~\bibnamefont{Klein}}, \bibinfo{journal}{Z.
  Phys.} \textbf{\bibinfo{volume}{37}}, \bibinfo{pages}{895}
  (\bibinfo{year}{1926}).

\bibitem[{\citenamefont{{Weinberg}}(1983)}]{Weinberg83}
\bibinfo{author}{\bibfnamefont{S.}~\bibnamefont{{Weinberg}}},
  \bibinfo{journal}{Physics Letters B} \textbf{\bibinfo{volume}{125}},
  \bibinfo{pages}{265} (\bibinfo{year}{1983}).

\bibitem[{\citenamefont{{Gleiser} and {Taylor}}(1985)}]{GT85}
\bibinfo{author}{\bibfnamefont{M.}~\bibnamefont{{Gleiser}}} \bibnamefont{and}
  \bibinfo{author}{\bibfnamefont{J.~G.} \bibnamefont{{Taylor}}},
  \bibinfo{journal}{\prd} \textbf{\bibinfo{volume}{31}}, \bibinfo{pages}{1904}
  (\bibinfo{year}{1985}).

\bibitem[{\citenamefont{{Overduin} and {Wesson}}(1997)}]{OW97}
\bibinfo{author}{\bibfnamefont{J.~M.} \bibnamefont{{Overduin}}}
  \bibnamefont{and} \bibinfo{author}{\bibfnamefont{P.~S.}
  \bibnamefont{{Wesson}}}, \bibinfo{journal}{Phys.Rep.}
  \textbf{\bibinfo{volume}{283}}, \bibinfo{pages}{303} (\bibinfo{year}{1997}).

\bibitem[{\citenamefont{Bekenstein}(1982)}]{Bekenstein82}
\bibinfo{author}{\bibfnamefont{J.~D.} \bibnamefont{Bekenstein}},
  \bibinfo{journal}{\prd} \textbf{\bibinfo{volume}{25}}, \bibinfo{pages}{1527}
  (\bibinfo{year}{1982}).

\bibitem[{\citenamefont{{Barrow} et~al.}(2002)\citenamefont{{Barrow},
  {Sandvik}, and {Magueijo}}}]{BSM02}
\bibinfo{author}{\bibfnamefont{J.~D.} \bibnamefont{{Barrow}}},
  \bibinfo{author}{\bibfnamefont{H.~B.} \bibnamefont{{Sandvik}}},
  \bibnamefont{and}
  \bibinfo{author}{\bibfnamefont{J.}~\bibnamefont{{Magueijo}}},
  \bibinfo{journal}{\prd} \textbf{\bibinfo{volume}{65}},
  \bibinfo{pages}{063504} (\bibinfo{year}{2002}).

\bibitem[{\citenamefont{{Barrow} and {Magueijo}}(2005)}]{BM05}
\bibinfo{author}{\bibfnamefont{J.~D.} \bibnamefont{{Barrow}}} \bibnamefont{and}
  \bibinfo{author}{\bibfnamefont{J.}~\bibnamefont{{Magueijo}}},
  \bibinfo{journal}{\prd} \textbf{\bibinfo{volume}{72}},
  \bibinfo{pages}{043521} (\bibinfo{year}{2005}),
  \eprint{arXiv:astro-ph/0503222}.

\bibitem[{\citenamefont{{Bize} et~al.}(2003)\citenamefont{{Bize}, {Diddams},
  {Tanaka}, {Tanner}, {Oskay}, {Drullinger}, {Parker}, {Heavner}, {Jefferts},
  {Hollberg} et~al.}}]{Bize03}
\bibinfo{author}{\bibfnamefont{S.}~\bibnamefont{{Bize}}},
  \bibinfo{author}{\bibfnamefont{S.~A.} \bibnamefont{{Diddams}}},
  \bibinfo{author}{\bibfnamefont{U.}~\bibnamefont{{Tanaka}}},
  \bibinfo{author}{\bibfnamefont{C.~E.} \bibnamefont{{Tanner}}},
  \bibinfo{author}{\bibfnamefont{W.~H.} \bibnamefont{{Oskay}}},
  \bibinfo{author}{\bibfnamefont{R.~E.} \bibnamefont{{Drullinger}}},
  \bibinfo{author}{\bibfnamefont{T.~E.} \bibnamefont{{Parker}}},
  \bibinfo{author}{\bibfnamefont{T.~P.} \bibnamefont{{Heavner}}},
  \bibinfo{author}{\bibfnamefont{S.~R.} \bibnamefont{{Jefferts}}},
  \bibinfo{author}{\bibfnamefont{L.}~\bibnamefont{{Hollberg}}},
  \bibnamefont{et~al.}, \bibinfo{journal}{\prl} \textbf{\bibinfo{volume}{90}},
  \bibinfo{pages}{150802} (\bibinfo{year}{2003}).

\bibitem[{\citenamefont{{Fischer} et~al.}(2004)\citenamefont{{Fischer},
  {Kolachevsky}, {Zimmermann}, {Holzwarth}, {Udem}, {H{\" a}nsch}, {Abgrall},
  {Gr{\" u}nert}, {Maksimovic}, {Bize} et~al.}}]{Fischer04}
\bibinfo{author}{\bibfnamefont{M.}~\bibnamefont{{Fischer}}},
  \bibinfo{author}{\bibfnamefont{N.}~\bibnamefont{{Kolachevsky}}},
  \bibinfo{author}{\bibfnamefont{M.}~\bibnamefont{{Zimmermann}}},
  \bibinfo{author}{\bibfnamefont{R.}~\bibnamefont{{Holzwarth}}},
  \bibinfo{author}{\bibfnamefont{T.}~\bibnamefont{{Udem}}},
  \bibinfo{author}{\bibfnamefont{T.~W.} \bibnamefont{{H{\" a}nsch}}},
  \bibinfo{author}{\bibfnamefont{M.}~\bibnamefont{{Abgrall}}},
  \bibinfo{author}{\bibfnamefont{J.}~\bibnamefont{{Gr{\" u}nert}}},
  \bibinfo{author}{\bibfnamefont{I.}~\bibnamefont{{Maksimovic}}},
  \bibinfo{author}{\bibfnamefont{S.}~\bibnamefont{{Bize}}},
  \bibnamefont{et~al.}, \bibinfo{journal}{\prl} \textbf{\bibinfo{volume}{92}},
  \bibinfo{pages}{230802} (\bibinfo{year}{2004}).

\bibitem[{\citenamefont{{Peik} et~al.}(2004)\citenamefont{{Peik}, {Lipphardt},
  {Schnatz}, {Schneider}, {Tamm}, and {Karshenboim}}}]{Peik04}
\bibinfo{author}{\bibfnamefont{E.}~\bibnamefont{{Peik}}},
  \bibinfo{author}{\bibfnamefont{B.}~\bibnamefont{{Lipphardt}}},
  \bibinfo{author}{\bibfnamefont{H.}~\bibnamefont{{Schnatz}}},
  \bibinfo{author}{\bibfnamefont{T.}~\bibnamefont{{Schneider}}},
  \bibinfo{author}{\bibfnamefont{C.}~\bibnamefont{{Tamm}}}, \bibnamefont{and}
  \bibinfo{author}{\bibfnamefont{S.~G.} \bibnamefont{{Karshenboim}}},
  \bibinfo{journal}{\prl} \textbf{\bibinfo{volume}{93}},
  \bibinfo{pages}{170801} (\bibinfo{year}{2004}), \eprint{physics/0402132}.

\bibitem[{\citenamefont{{Prestage} et~al.}(1995)\citenamefont{{Prestage},
  {Tjoelker}, and {Maleki}}}]{PTM95}
\bibinfo{author}{\bibfnamefont{J.~D.} \bibnamefont{{Prestage}}},
  \bibinfo{author}{\bibfnamefont{R.~L.} \bibnamefont{{Tjoelker}}},
  \bibnamefont{and} \bibinfo{author}{\bibfnamefont{L.}~\bibnamefont{{Maleki}}},
  \bibinfo{journal}{\prl} \textbf{\bibinfo{volume}{74}}, \bibinfo{pages}{3511}
  (\bibinfo{year}{1995}).

\bibitem[{\citenamefont{{Sortais} et~al.}(2000)\citenamefont{{Sortais}, {Bize},
  {Abgrall}, {Zhang}, {Nicolas}, {Mandache}, P, {Laurent}, {Santarelli},
  {Dimarcq} et~al.}}]{Sortais00}
\bibinfo{author}{\bibfnamefont{Y.}~\bibnamefont{{Sortais}}},
  \bibinfo{author}{\bibfnamefont{S.}~\bibnamefont{{Bize}}},
  \bibinfo{author}{\bibfnamefont{M.}~\bibnamefont{{Abgrall}}},
  \bibinfo{author}{\bibfnamefont{S.}~\bibnamefont{{Zhang}}},
  \bibinfo{author}{\bibnamefont{{Nicolas}}},
  \bibinfo{author}{\bibfnamefont{C.}~\bibnamefont{{Mandache}}},
  \bibinfo{author}{\bibfnamefont{L.}~\bibnamefont{P}},
  \bibinfo{author}{\bibfnamefont{P.}~\bibnamefont{{Laurent}}},
  \bibinfo{author}{\bibfnamefont{G.}~\bibnamefont{{Santarelli}}},
  \bibinfo{author}{\bibfnamefont{N.}~\bibnamefont{{Dimarcq}}},
  \bibnamefont{et~al.}, \bibinfo{journal}{Physica Scripta}
  \textbf{\bibinfo{volume}{T95}}, \bibinfo{pages}{50} (\bibinfo{year}{2000}).

\bibitem[{\citenamefont{{Marion} et~al.}(2003)\citenamefont{{Marion}, {Pereira
  Dos Santos}, {Abgrall}, {Zhang}, {Sortais}, {Bize}, {Maksimovic}, {Calonico},
  {Gr{\" u}nert}, {Mandache} et~al.}}]{Marion03}
\bibinfo{author}{\bibfnamefont{H.}~\bibnamefont{{Marion}}},
  \bibinfo{author}{\bibfnamefont{F.}~\bibnamefont{{Pereira Dos Santos}}},
  \bibinfo{author}{\bibfnamefont{M.}~\bibnamefont{{Abgrall}}},
  \bibinfo{author}{\bibfnamefont{S.}~\bibnamefont{{Zhang}}},
  \bibinfo{author}{\bibfnamefont{Y.}~\bibnamefont{{Sortais}}},
  \bibinfo{author}{\bibfnamefont{S.}~\bibnamefont{{Bize}}},
  \bibinfo{author}{\bibfnamefont{I.}~\bibnamefont{{Maksimovic}}},
  \bibinfo{author}{\bibfnamefont{D.}~\bibnamefont{{Calonico}}},
  \bibinfo{author}{\bibfnamefont{J.}~\bibnamefont{{Gr{\" u}nert}}},
  \bibinfo{author}{\bibfnamefont{C.}~\bibnamefont{{Mandache}}},
  \bibnamefont{et~al.}, \bibinfo{journal}{\prl} \textbf{\bibinfo{volume}{90}},
  \bibinfo{pages}{150801} (\bibinfo{year}{2003}).

\bibitem[{\citenamefont{{Fujii} et~al.}(2000)\citenamefont{{Fujii}, {Iwamoto},
  {Fukahori}, {Ohnuki}, {Nakagawa}, {Hidaka}, {Oura}, and {M{\"
  o}ller}}}]{Fujii00}
\bibinfo{author}{\bibfnamefont{Y.}~\bibnamefont{{Fujii}}},
  \bibinfo{author}{\bibfnamefont{A.}~\bibnamefont{{Iwamoto}}},
  \bibinfo{author}{\bibfnamefont{T.}~\bibnamefont{{Fukahori}}},
  \bibinfo{author}{\bibfnamefont{T.}~\bibnamefont{{Ohnuki}}},
  \bibinfo{author}{\bibfnamefont{M.}~\bibnamefont{{Nakagawa}}},
  \bibinfo{author}{\bibfnamefont{H.}~\bibnamefont{{Hidaka}}},
  \bibinfo{author}{\bibfnamefont{Y.}~\bibnamefont{{Oura}}}, \bibnamefont{and}
  \bibinfo{author}{\bibfnamefont{P.}~\bibnamefont{{M{\" o}ller}}},
  \bibinfo{journal}{Nuclear Physics B} \textbf{\bibinfo{volume}{573}},
  \bibinfo{pages}{377} (\bibinfo{year}{2000}).

\bibitem[{\citenamefont{{Damour} and {Dyson}}(1996)}]{DD96}
\bibinfo{author}{\bibfnamefont{T.}~\bibnamefont{{Damour}}} \bibnamefont{and}
  \bibinfo{author}{\bibfnamefont{F.}~\bibnamefont{{Dyson}}},
  \bibinfo{journal}{Nuclear Physics B} \textbf{\bibinfo{volume}{480}},
  \bibinfo{pages}{37} (\bibinfo{year}{1996}).

\bibitem[{\citenamefont{{Olive} et~al.}(2004)\citenamefont{{Olive}, {Pospelov},
  {Qian}, {Manh{\`e}s}, {Vangioni-Flam}, {Coc}, and {Cass{\'e}}}}]{Olive04b}
\bibinfo{author}{\bibfnamefont{K.~A.} \bibnamefont{{Olive}}},
  \bibinfo{author}{\bibfnamefont{M.}~\bibnamefont{{Pospelov}}},
  \bibinfo{author}{\bibfnamefont{Y.~Z.} \bibnamefont{{Qian}}},
  \bibinfo{author}{\bibfnamefont{G.}~\bibnamefont{{Manh{\`e}s}}},
  \bibinfo{author}{\bibfnamefont{E.}~\bibnamefont{{Vangioni-Flam}}},
  \bibinfo{author}{\bibfnamefont{A.}~\bibnamefont{{Coc}}}, \bibnamefont{and}
  \bibinfo{author}{\bibfnamefont{M.}~\bibnamefont{{Cass{\'e}}}},
  \bibinfo{journal}{\prd} \textbf{\bibinfo{volume}{69}},
  \bibinfo{pages}{027701} (\bibinfo{year}{2004}), \eprint{astro-ph/0309252}.

\bibitem[{\citenamefont{{Webb} et~al.}(1999)\citenamefont{{Webb}, {Flambaum},
  {Churchill}, {Drinkwater}, and {Barrow}}}]{Webb99}
\bibinfo{author}{\bibfnamefont{J.~K.} \bibnamefont{{Webb}}},
  \bibinfo{author}{\bibfnamefont{V.~V.} \bibnamefont{{Flambaum}}},
  \bibinfo{author}{\bibfnamefont{C.~W.} \bibnamefont{{Churchill}}},
  \bibinfo{author}{\bibfnamefont{M.~J.} \bibnamefont{{Drinkwater}}},
  \bibnamefont{and} \bibinfo{author}{\bibfnamefont{J.~D.}
  \bibnamefont{{Barrow}}}, \bibinfo{journal}{\prl}
  \textbf{\bibinfo{volume}{82}}, \bibinfo{pages}{884} (\bibinfo{year}{1999}).

\bibitem[{\citenamefont{{Webb} et~al.}(2001)\citenamefont{{Webb}, {Murphy},
  {Flambaum}, {Dzuba}, {Barrow}, {Churchill}, {Prochaska}, and
  {Wolfe}}}]{Webb01}
\bibinfo{author}{\bibfnamefont{J.~K.} \bibnamefont{{Webb}}},
  \bibinfo{author}{\bibfnamefont{M.~T.} \bibnamefont{{Murphy}}},
  \bibinfo{author}{\bibfnamefont{V.~V.} \bibnamefont{{Flambaum}}},
  \bibinfo{author}{\bibfnamefont{V.~A.} \bibnamefont{{Dzuba}}},
  \bibinfo{author}{\bibfnamefont{J.~D.} \bibnamefont{{Barrow}}},
  \bibinfo{author}{\bibfnamefont{C.~W.} \bibnamefont{{Churchill}}},
  \bibinfo{author}{\bibfnamefont{J.~X.} \bibnamefont{{Prochaska}}},
  \bibnamefont{and} \bibinfo{author}{\bibfnamefont{A.~M.}
  \bibnamefont{{Wolfe}}}, \bibinfo{journal}{\prl}
  \textbf{\bibinfo{volume}{87}}, \bibinfo{pages}{091301}
  (\bibinfo{year}{2001}).

\bibitem[{\citenamefont{{Murphy}
  et~al.}(2001{\natexlab{a}})\citenamefont{{Murphy}, {Webb}, {Flambaum},
  {Dzuba}, {Churchill}, {Prochaska}, {Barrow}, and {Wolfe}}}]{Murphy01a}
\bibinfo{author}{\bibfnamefont{M.~T.} \bibnamefont{{Murphy}}},
  \bibinfo{author}{\bibfnamefont{J.~K.} \bibnamefont{{Webb}}},
  \bibinfo{author}{\bibfnamefont{V.~V.} \bibnamefont{{Flambaum}}},
  \bibinfo{author}{\bibfnamefont{V.~A.} \bibnamefont{{Dzuba}}},
  \bibinfo{author}{\bibfnamefont{C.~W.} \bibnamefont{{Churchill}}},
  \bibinfo{author}{\bibfnamefont{J.~X.} \bibnamefont{{Prochaska}}},
  \bibinfo{author}{\bibfnamefont{J.~D.} \bibnamefont{{Barrow}}},
  \bibnamefont{and} \bibinfo{author}{\bibfnamefont{A.~M.}
  \bibnamefont{{Wolfe}}}, \bibinfo{journal}{Mon.Not.R.Astron.Soc.}
  \textbf{\bibinfo{volume}{327}}, \bibinfo{pages}{1208}
  (\bibinfo{year}{2001}{\natexlab{a}}).

\bibitem[{\citenamefont{{Murphy}
  et~al.}(2001{\natexlab{b}})\citenamefont{{Murphy}, {Webb}, {Flambaum},
  {Prochaska}, and {Wolfe}}}]{Murphy01b}
\bibinfo{author}{\bibfnamefont{M.~T.} \bibnamefont{{Murphy}}},
  \bibinfo{author}{\bibfnamefont{J.~K.} \bibnamefont{{Webb}}},
  \bibinfo{author}{\bibfnamefont{V.~V.} \bibnamefont{{Flambaum}}},
  \bibinfo{author}{\bibfnamefont{J.~X.} \bibnamefont{{Prochaska}}},
  \bibnamefont{and} \bibinfo{author}{\bibfnamefont{A.~M.}
  \bibnamefont{{Wolfe}}}, \bibinfo{journal}{Mon.Not.R.Astron.Soc.}
  \textbf{\bibinfo{volume}{327}}, \bibinfo{pages}{1237}
  (\bibinfo{year}{2001}{\natexlab{b}}).

\bibitem[{\citenamefont{{Murphy} et~al.}(2003)\citenamefont{{Murphy}, {Webb},
  and {Flambaum}}}]{Murphy03b}
\bibinfo{author}{\bibfnamefont{M.~T.} \bibnamefont{{Murphy}}},
  \bibinfo{author}{\bibfnamefont{J.~K.} \bibnamefont{{Webb}}},
  \bibnamefont{and} \bibinfo{author}{\bibfnamefont{V.~V.}
  \bibnamefont{{Flambaum}}}, \bibinfo{journal}{Mon.Not.R.Astron.Soc.}
  \textbf{\bibinfo{volume}{345}}, \bibinfo{pages}{609} (\bibinfo{year}{2003}).

\bibitem[{\citenamefont{{Ivanchik} et~al.}(2005)\citenamefont{{Ivanchik},
  {Petitjean}, {Varshalovich}, {Aracil}, {Srianand}, {Chand}, {Ledoux}, and
  {Boiss{\'e}}}}]{Ivanchik05}
\bibinfo{author}{\bibfnamefont{A.}~\bibnamefont{{Ivanchik}}},
  \bibinfo{author}{\bibfnamefont{P.}~\bibnamefont{{Petitjean}}},
  \bibinfo{author}{\bibfnamefont{D.}~\bibnamefont{{Varshalovich}}},
  \bibinfo{author}{\bibfnamefont{B.}~\bibnamefont{{Aracil}}},
  \bibinfo{author}{\bibfnamefont{R.}~\bibnamefont{{Srianand}}},
  \bibinfo{author}{\bibfnamefont{H.}~\bibnamefont{{Chand}}},
  \bibinfo{author}{\bibfnamefont{C.}~\bibnamefont{{Ledoux}}}, \bibnamefont{and}
  \bibinfo{author}{\bibfnamefont{P.}~\bibnamefont{{Boiss{\'e}}}},
  \bibinfo{journal}{Astron. and Astrophys.} \textbf{\bibinfo{volume}{440}},
  \bibinfo{pages}{45} (\bibinfo{year}{2005}), \eprint{astro-ph/0507174}.

\bibitem[{\citenamefont{{Tzanavaris} et~al.}(2007)\citenamefont{{Tzanavaris},
  {Murphy}, {Webb}, {Flambaum}, and {Curran}}}]{Tzana07}
\bibinfo{author}{\bibfnamefont{P.}~\bibnamefont{{Tzanavaris}}},
  \bibinfo{author}{\bibfnamefont{M.~T.} \bibnamefont{{Murphy}}},
  \bibinfo{author}{\bibfnamefont{J.~K.} \bibnamefont{{Webb}}},
  \bibinfo{author}{\bibfnamefont{V.~V.} \bibnamefont{{Flambaum}}},
  \bibnamefont{and} \bibinfo{author}{\bibfnamefont{S.~J.}
  \bibnamefont{{Curran}}}, \bibinfo{journal}{Mon.Not.Roy.Astron.Soc.}
  \textbf{\bibinfo{volume}{374}}, \bibinfo{pages}{634} (\bibinfo{year}{2007}),
  \eprint{astro-ph/0610326}.

\bibitem[{\citenamefont{{Mart{\'{\i}}nez Fiorenzano}
  et~al.}(2003)\citenamefont{{Mart{\'{\i}}nez Fiorenzano}, {Vladilo}, and
  {Bonifacio}}}]{MVB04}
\bibinfo{author}{\bibfnamefont{A.~F.} \bibnamefont{{Mart{\'{\i}}nez
  Fiorenzano}}}, \bibinfo{author}{\bibfnamefont{G.}~\bibnamefont{{Vladilo}}},
  \bibnamefont{and}
  \bibinfo{author}{\bibfnamefont{P.}~\bibnamefont{{Bonifacio}}},
  \bibinfo{journal}{Societa Astronomica Italiana Memorie Supplement}
  \textbf{\bibinfo{volume}{3}}, \bibinfo{pages}{252} (\bibinfo{year}{2003}).

\bibitem[{\citenamefont{{Quast} et~al.}(2004)\citenamefont{{Quast}, {Reimers},
  and {Levshakov}}}]{QRL04}
\bibinfo{author}{\bibfnamefont{R.}~\bibnamefont{{Quast}}},
  \bibinfo{author}{\bibfnamefont{D.}~\bibnamefont{{Reimers}}},
  \bibnamefont{and} \bibinfo{author}{\bibfnamefont{S.~A.}
  \bibnamefont{{Levshakov}}}, \bibinfo{journal}{Astron.Astrophys.}
  \textbf{\bibinfo{volume}{415}}, \bibinfo{pages}{L7} (\bibinfo{year}{2004}).

\bibitem[{\citenamefont{{Bahcall} et~al.}(2004)\citenamefont{{Bahcall},
  {Steinhardt}, and {Schlegel}}}]{Bahcall04}
\bibinfo{author}{\bibfnamefont{J.~N.} \bibnamefont{{Bahcall}}},
  \bibinfo{author}{\bibfnamefont{C.~L.} \bibnamefont{{Steinhardt}}},
  \bibnamefont{and}
  \bibinfo{author}{\bibfnamefont{D.}~\bibnamefont{{Schlegel}}},
  \bibinfo{journal}{Astrophys.J.} \textbf{\bibinfo{volume}{600}},
  \bibinfo{pages}{520} (\bibinfo{year}{2004}).

\bibitem[{\citenamefont{{Srianand} et~al.}(2004)\citenamefont{{Srianand},
  {Chand}, {Petitjean}, and {Aracil}}}]{Srianand04}
\bibinfo{author}{\bibfnamefont{R.}~\bibnamefont{{Srianand}}},
  \bibinfo{author}{\bibfnamefont{H.}~\bibnamefont{{Chand}}},
  \bibinfo{author}{\bibfnamefont{P.}~\bibnamefont{{Petitjean}}},
  \bibnamefont{and} \bibinfo{author}{\bibfnamefont{B.}~\bibnamefont{{Aracil}}},
  \bibinfo{journal}{\prl} \textbf{\bibinfo{volume}{92}},
  \bibinfo{pages}{121302} (\bibinfo{year}{2004}).

\bibitem[{\citenamefont{{Mosquera} et~al.}(2008)\citenamefont{{Mosquera},
  {Sc{\'o}ccola}, {Landau}, and {Vucetich}}}]{Mosquera07}
\bibinfo{author}{\bibfnamefont{M.~E.} \bibnamefont{{Mosquera}}},
  \bibinfo{author}{\bibfnamefont{C.~G.} \bibnamefont{{Sc{\'o}ccola}}},
  \bibinfo{author}{\bibfnamefont{S.~J.} \bibnamefont{{Landau}}},
  \bibnamefont{and}
  \bibinfo{author}{\bibfnamefont{H.}~\bibnamefont{{Vucetich}}},
  \bibinfo{journal}{Astronomy and Astrophysics} \textbf{\bibinfo{volume}{478}},
  \bibinfo{pages}{675} (\bibinfo{year}{2008}), \eprint{arXiv:0707.0661}.

\bibitem[{\citenamefont{{Sc{\'o}ccola}
  et~al.}(2008)\citenamefont{{Sc{\'o}ccola}, {Mosquera}, {Landau}, and
  {Vucetich}}}]{Scoccola07}
\bibinfo{author}{\bibfnamefont{C.~G.} \bibnamefont{{Sc{\'o}ccola}}},
  \bibinfo{author}{\bibfnamefont{M.~E.} \bibnamefont{{Mosquera}}},
  \bibinfo{author}{\bibfnamefont{S.~J.} \bibnamefont{{Landau}}},
  \bibnamefont{and}
  \bibinfo{author}{\bibfnamefont{H.}~\bibnamefont{{Vucetich}}},
  \bibinfo{journal}{\apj} \textbf{\bibinfo{volume}{681}}, \bibinfo{pages}{737}
  (\bibinfo{year}{2008}), \eprint{arXiv:0803.0247}.

\bibitem[{\citenamefont{{Bergstr{\" o}m} et~al.}(1999)\citenamefont{{Bergstr{\"
  o}m}, {Iguri}, and {Rubinstein}}}]{Iguri99}
\bibinfo{author}{\bibfnamefont{L.}~\bibnamefont{{Bergstr{\" o}m}}},
  \bibinfo{author}{\bibfnamefont{S.}~\bibnamefont{{Iguri}}}, \bibnamefont{and}
  \bibinfo{author}{\bibfnamefont{H.}~\bibnamefont{{Rubinstein}}},
  \bibinfo{journal}{\prd} \textbf{\bibinfo{volume}{60}}, \bibinfo{pages}{45005}
  (\bibinfo{year}{1999}).

\bibitem[{\citenamefont{{Nollett} and {Lopez}}(2002)}]{Nollet}
\bibinfo{author}{\bibfnamefont{K.~M.} \bibnamefont{{Nollett}}}
  \bibnamefont{and} \bibinfo{author}{\bibfnamefont{R.~E.}
  \bibnamefont{{Lopez}}}, \bibinfo{journal}{Phys. Rev.}
  \textbf{\bibinfo{volume}{D66}}, \bibinfo{pages}{063507}
  (\bibinfo{year}{2002}).

\bibitem[{\citenamefont{{Yoo} and {Scherrer}}(2003)}]{YS03}
\bibinfo{author}{\bibfnamefont{J.~J.} \bibnamefont{{Yoo}}} \bibnamefont{and}
  \bibinfo{author}{\bibfnamefont{R.~J.} \bibnamefont{{Scherrer}}},
  \bibinfo{journal}{\prd} \textbf{\bibinfo{volume}{67}},
  \bibinfo{pages}{043517} (\bibinfo{year}{2003}).

\bibitem[{\citenamefont{{Landau} et~al.}(2006)\citenamefont{{Landau},
  {Mosquera}, and {Vucetich}}}]{LMV06}
\bibinfo{author}{\bibfnamefont{S.~J.} \bibnamefont{{Landau}}},
  \bibinfo{author}{\bibfnamefont{M.~E.} \bibnamefont{{Mosquera}}},
  \bibnamefont{and}
  \bibinfo{author}{\bibfnamefont{H.}~\bibnamefont{{Vucetich}}},
  \bibinfo{journal}{Astrophys. J.} \textbf{\bibinfo{volume}{637}},
  \bibinfo{pages}{38} (\bibinfo{year}{2006}).

\bibitem[{\citenamefont{{Chamoun} et~al.}(2007)\citenamefont{{Chamoun},
  {Landau}, {Mosquera}, and {Vucetich}}}]{Chamoun07}
\bibinfo{author}{\bibfnamefont{N.}~\bibnamefont{{Chamoun}}},
  \bibinfo{author}{\bibfnamefont{S.~J.} \bibnamefont{{Landau}}},
  \bibinfo{author}{\bibfnamefont{M.~E.} \bibnamefont{{Mosquera}}},
  \bibnamefont{and}
  \bibinfo{author}{\bibfnamefont{H.}~\bibnamefont{{Vucetich}}},
  \bibinfo{journal}{Journal of Physics G Nuclear Physics}
  \textbf{\bibinfo{volume}{34}}, \bibinfo{pages}{163} (\bibinfo{year}{2007}),
  \eprint{astro-ph/0508378}.

\bibitem[{\citenamefont{{Campbell} and {Olive}}(1995)}]{CO95}
\bibinfo{author}{\bibfnamefont{B.~A.} \bibnamefont{{Campbell}}}
  \bibnamefont{and} \bibinfo{author}{\bibfnamefont{K.~A.}
  \bibnamefont{{Olive}}}, \bibinfo{journal}{Physics Letters B}
  \textbf{\bibinfo{volume}{345}}, \bibinfo{pages}{429} (\bibinfo{year}{1995}).

\bibitem[{\citenamefont{{Ichikawa} and {Kawasaki}}(2002)}]{Ichi02}
\bibinfo{author}{\bibfnamefont{K.}~\bibnamefont{{Ichikawa}}} \bibnamefont{and}
  \bibinfo{author}{\bibfnamefont{M.}~\bibnamefont{{Kawasaki}}},
  \bibinfo{journal}{Phys. Rev.} \textbf{\bibinfo{volume}{D65}},
  \bibinfo{pages}{123511} (\bibinfo{year}{2002}).

\bibitem[{\citenamefont{{Ichikawa} and {Kawasaki}}(2004)}]{ichi04}
\bibinfo{author}{\bibfnamefont{K.}~\bibnamefont{{Ichikawa}}} \bibnamefont{and}
  \bibinfo{author}{\bibfnamefont{M.}~\bibnamefont{{Kawasaki}}},
  \bibinfo{journal}{\prd} \textbf{\bibinfo{volume}{69}},
  \bibinfo{pages}{123506} (\bibinfo{year}{2004}).

\bibitem[{\citenamefont{{M{\" u}ller} et~al.}(2004)\citenamefont{{M{\" u}ller},
  {Sch{\" a}fer}, and {Wetterich}}}]{mueller04}
\bibinfo{author}{\bibfnamefont{C.~M.} \bibnamefont{{M{\" u}ller}}},
  \bibinfo{author}{\bibfnamefont{G.}~\bibnamefont{{Sch{\" a}fer}}},
  \bibnamefont{and}
  \bibinfo{author}{\bibfnamefont{C.}~\bibnamefont{{Wetterich}}},
  \bibinfo{journal}{Phys.Rev.D} \textbf{\bibinfo{volume}{70}},
  \bibinfo{pages}{083504} (\bibinfo{year}{2004}).

\bibitem[{\citenamefont{{Coc} et~al.}(2007)\citenamefont{{Coc}, {Nunes},
  {Olive}, {Uzan}, and {Vangioni}}}]{coc07}
\bibinfo{author}{\bibfnamefont{A.}~\bibnamefont{{Coc}}},
  \bibinfo{author}{\bibfnamefont{N.~J.} \bibnamefont{{Nunes}}},
  \bibinfo{author}{\bibfnamefont{K.~A.} \bibnamefont{{Olive}}},
  \bibinfo{author}{\bibfnamefont{J.-P.} \bibnamefont{{Uzan}}},
  \bibnamefont{and}
  \bibinfo{author}{\bibfnamefont{E.}~\bibnamefont{{Vangioni}}},
  \bibinfo{journal}{\prd} \textbf{\bibinfo{volume}{76}},
  \bibinfo{pages}{023511} (\bibinfo{year}{2007}),
  \eprint{arXiv:astro-ph/0610733}.

\bibitem[{\citenamefont{{Cyburt} et~al.}(2005)\citenamefont{{Cyburt}, {Fields},
  {Olive}, and {Skillman}}}]{cyburt05}
\bibinfo{author}{\bibfnamefont{R.~H.} \bibnamefont{{Cyburt}}},
  \bibinfo{author}{\bibfnamefont{B.~D.} \bibnamefont{{Fields}}},
  \bibinfo{author}{\bibfnamefont{K.~A.} \bibnamefont{{Olive}}},
  \bibnamefont{and}
  \bibinfo{author}{\bibfnamefont{E.}~\bibnamefont{{Skillman}}},
  \bibinfo{journal}{Astroparticle Physics} \textbf{\bibinfo{volume}{23}},
  \bibinfo{pages}{313} (\bibinfo{year}{2005}), \eprint{arXiv:astro-ph/0408033}.

\bibitem[{\citenamefont{{Dent} et~al.}(2007)\citenamefont{{Dent}, {Stern}, and
  {Wetterich}}}]{Dent07}
\bibinfo{author}{\bibfnamefont{T.}~\bibnamefont{{Dent}}},
  \bibinfo{author}{\bibfnamefont{S.}~\bibnamefont{{Stern}}}, \bibnamefont{and}
  \bibinfo{author}{\bibfnamefont{C.}~\bibnamefont{{Wetterich}}},
  \bibinfo{journal}{\prd} \textbf{\bibinfo{volume}{76}},
  \bibinfo{pages}{063513} (\bibinfo{year}{2007}), \eprint{arXiv:0705.0696}.

\bibitem[{\citenamefont{{Flambaum} and {Shuryak}}(2002)}]{Flambaum02}
\bibinfo{author}{\bibfnamefont{V.~V.} \bibnamefont{{Flambaum}}}
  \bibnamefont{and} \bibinfo{author}{\bibfnamefont{E.~V.}
  \bibnamefont{{Shuryak}}}, \bibinfo{journal}{\prd}
  \textbf{\bibinfo{volume}{65}}, \bibinfo{pages}{103503}
  (\bibinfo{year}{2002}).

\bibitem[{\citenamefont{{Flambaum} et~al.}(2004)\citenamefont{{Flambaum},
  {Leinweber}, {Thomas}, and {Young}}}]{Flambaum04b}
\bibinfo{author}{\bibfnamefont{V.~V.} \bibnamefont{{Flambaum}}},
  \bibinfo{author}{\bibfnamefont{D.~B.} \bibnamefont{{Leinweber}}},
  \bibinfo{author}{\bibfnamefont{A.~W.} \bibnamefont{{Thomas}}},
  \bibnamefont{and} \bibinfo{author}{\bibfnamefont{R.~D.}
  \bibnamefont{{Young}}}, \bibinfo{journal}{\prd}
  \textbf{\bibinfo{volume}{69}}, \bibinfo{pages}{115006}
  (\bibinfo{year}{2004}).

\bibitem[{\citenamefont{{Kneller} and {McLaughlin}}(2003)}]{KM03}
\bibinfo{author}{\bibfnamefont{J.~P.} \bibnamefont{{Kneller}}}
  \bibnamefont{and} \bibinfo{author}{\bibfnamefont{G.~C.}
  \bibnamefont{{McLaughlin}}}, \bibinfo{journal}{\prd}
  \textbf{\bibinfo{volume}{68}}, \bibinfo{pages}{103508}
  (\bibinfo{year}{2003}).

\bibitem[{\citenamefont{{Martins} et~al.}(2002)\citenamefont{{Martins},
  {Melchiorri}, {Trotta}, {Bean}, {Rocha}, {Avelino}, and {Viana}}}]{Martins02}
\bibinfo{author}{\bibfnamefont{C.~J.~A.~P.} \bibnamefont{{Martins}}},
  \bibinfo{author}{\bibfnamefont{A.}~\bibnamefont{{Melchiorri}}},
  \bibinfo{author}{\bibfnamefont{R.}~\bibnamefont{{Trotta}}},
  \bibinfo{author}{\bibfnamefont{R.}~\bibnamefont{{Bean}}},
  \bibinfo{author}{\bibfnamefont{G.}~\bibnamefont{{Rocha}}},
  \bibinfo{author}{\bibfnamefont{P.~P.} \bibnamefont{{Avelino}}},
  \bibnamefont{and} \bibinfo{author}{\bibfnamefont{P.~T.~P.}
  \bibnamefont{{Viana}}}, \bibinfo{journal}{\prd}
  \textbf{\bibinfo{volume}{66}}, \bibinfo{pages}{023505}
  (\bibinfo{year}{2002}).

\bibitem[{\citenamefont{{Rocha} et~al.}(2003)\citenamefont{{Rocha}, {Trotta},
  {Martins}, {Melchiorri}, {Avelino}, and {Viana}}}]{Rocha03}
\bibinfo{author}{\bibfnamefont{G.}~\bibnamefont{{Rocha}}},
  \bibinfo{author}{\bibfnamefont{R.}~\bibnamefont{{Trotta}}},
  \bibinfo{author}{\bibfnamefont{C.~J.~A.~P.} \bibnamefont{{Martins}}},
  \bibinfo{author}{\bibfnamefont{A.}~\bibnamefont{{Melchiorri}}},
  \bibinfo{author}{\bibfnamefont{P.~P.} \bibnamefont{{Avelino}}},
  \bibnamefont{and} \bibinfo{author}{\bibfnamefont{P.~T.~P.}
  \bibnamefont{{Viana}}}, \bibinfo{journal}{New Astronomy Review}
  \textbf{\bibinfo{volume}{47}}, \bibinfo{pages}{863} (\bibinfo{year}{2003}).

\bibitem[{\citenamefont{{Ichikawa} et~al.}(2006)\citenamefont{{Ichikawa},
  {Kanzaki}, and {Kawasaki}}}]{ichi06}
\bibinfo{author}{\bibfnamefont{K.}~\bibnamefont{{Ichikawa}}},
  \bibinfo{author}{\bibfnamefont{T.}~\bibnamefont{{Kanzaki}}},
  \bibnamefont{and}
  \bibinfo{author}{\bibfnamefont{M.}~\bibnamefont{{Kawasaki}}},
  \bibinfo{journal}{\prd} \textbf{\bibinfo{volume}{74}},
  \bibinfo{pages}{023515} (\bibinfo{year}{2006}), \eprint{astro-ph/0602577}.

\bibitem[{\citenamefont{{Spergel} et~al.}(2003)\citenamefont{{Spergel},
  {Verde}, {Peiris}, {Komatsu}, {Nolta}, {Bennett}, {Halpern}, {Hinshaw},
  {Jarosik}, {Kogut} et~al.}}]{wmapest}
\bibinfo{author}{\bibfnamefont{D.~N.} \bibnamefont{{Spergel}}},
  \bibinfo{author}{\bibfnamefont{L.}~\bibnamefont{{Verde}}},
  \bibinfo{author}{\bibfnamefont{H.~V.} \bibnamefont{{Peiris}}},
  \bibinfo{author}{\bibfnamefont{E.}~\bibnamefont{{Komatsu}}},
  \bibinfo{author}{\bibfnamefont{M.~R.} \bibnamefont{{Nolta}}},
  \bibinfo{author}{\bibfnamefont{C.~L.} \bibnamefont{{Bennett}}},
  \bibinfo{author}{\bibfnamefont{M.}~\bibnamefont{{Halpern}}},
  \bibinfo{author}{\bibfnamefont{G.}~\bibnamefont{{Hinshaw}}},
  \bibinfo{author}{\bibfnamefont{N.}~\bibnamefont{{Jarosik}}},
  \bibinfo{author}{\bibfnamefont{A.}~\bibnamefont{{Kogut}}},
  \bibnamefont{et~al.}, \bibinfo{journal}{Astrophys.J.Suppl.Ser.}
  \textbf{\bibinfo{volume}{148}}, \bibinfo{pages}{175} (\bibinfo{year}{2003}).

\bibitem[{\citenamefont{{Spergel} et~al.}(2007)\citenamefont{{Spergel}, {Bean},
  {Dor{\'e}}, {Nolta}, {Bennett}, {Dunkley}, {Hinshaw}, {Jarosik}, {Komatsu},
  {Page} et~al.}}]{wmap3}
\bibinfo{author}{\bibfnamefont{D.~N.} \bibnamefont{{Spergel}}},
  \bibinfo{author}{\bibfnamefont{R.}~\bibnamefont{{Bean}}},
  \bibinfo{author}{\bibfnamefont{O.}~\bibnamefont{{Dor{\'e}}}},
  \bibinfo{author}{\bibfnamefont{M.~R.} \bibnamefont{{Nolta}}},
  \bibinfo{author}{\bibfnamefont{C.~L.} \bibnamefont{{Bennett}}},
  \bibinfo{author}{\bibfnamefont{J.}~\bibnamefont{{Dunkley}}},
  \bibinfo{author}{\bibfnamefont{G.}~\bibnamefont{{Hinshaw}}},
  \bibinfo{author}{\bibfnamefont{N.}~\bibnamefont{{Jarosik}}},
  \bibinfo{author}{\bibfnamefont{E.}~\bibnamefont{{Komatsu}}},
  \bibinfo{author}{\bibfnamefont{L.}~\bibnamefont{{Page}}},
  \bibnamefont{et~al.}, \bibinfo{journal}{Astrophys.J.Suppl.Ser.}
  \textbf{\bibinfo{volume}{170}}, \bibinfo{pages}{377} (\bibinfo{year}{2007}),
  \eprint{arXiv:astro-ph/0603449}.

\bibitem[{\citenamefont{Sanchez et~al.}(2006)\citenamefont{Sanchez, Baugh,
  Percival, Peacock, Padilla, Cole, Frenk, and Norberg}}]{Sanchez06}
\bibinfo{author}{\bibfnamefont{A.~G.} \bibnamefont{Sanchez}},
  \bibinfo{author}{\bibfnamefont{C.~M.} \bibnamefont{Baugh}},
  \bibinfo{author}{\bibfnamefont{W.~J.} \bibnamefont{Percival}},
  \bibinfo{author}{\bibfnamefont{J.~A.} \bibnamefont{Peacock}},
  \bibinfo{author}{\bibfnamefont{N.~D.} \bibnamefont{Padilla}},
  \bibinfo{author}{\bibfnamefont{S.}~\bibnamefont{Cole}},
  \bibinfo{author}{\bibfnamefont{C.~S.} \bibnamefont{Frenk}}, \bibnamefont{and}
  \bibinfo{author}{\bibfnamefont{P.}~\bibnamefont{Norberg}},
  \bibinfo{journal}{Mon.Not.Roy.Astron.Soc.} \textbf{\bibinfo{volume}{366}},
  \bibinfo{pages}{189} (\bibinfo{year}{2006}).

\bibitem[{\citenamefont{{Pettini} and {Bowen}}(2001)}]{pettini}
\bibinfo{author}{\bibfnamefont{M.}~\bibnamefont{{Pettini}}} \bibnamefont{and}
  \bibinfo{author}{\bibfnamefont{D.~V.} \bibnamefont{{Bowen}}},
  \bibinfo{journal}{Astrophys.J.} \textbf{\bibinfo{volume}{560}},
  \bibinfo{pages}{41} (\bibinfo{year}{2001}).

\bibitem[{\citenamefont{{O'Meara} et~al.}(2001)\citenamefont{{O'Meara},
  {Tytler}, {Kirkman}, {Suzuki}, {Prochaska}, {Lubin}, and {Wolfe}}}]{omeara}
\bibinfo{author}{\bibfnamefont{J.~M.} \bibnamefont{{O'Meara}}},
  \bibinfo{author}{\bibfnamefont{D.}~\bibnamefont{{Tytler}}},
  \bibinfo{author}{\bibfnamefont{D.}~\bibnamefont{{Kirkman}}},
  \bibinfo{author}{\bibfnamefont{N.}~\bibnamefont{{Suzuki}}},
  \bibinfo{author}{\bibfnamefont{J.~X.} \bibnamefont{{Prochaska}}},
  \bibinfo{author}{\bibfnamefont{D.}~\bibnamefont{{Lubin}}}, \bibnamefont{and}
  \bibinfo{author}{\bibfnamefont{A.~M.} \bibnamefont{{Wolfe}}},
  \bibinfo{journal}{Astrophys.J.} \textbf{\bibinfo{volume}{552}},
  \bibinfo{pages}{718} (\bibinfo{year}{2001}).

\bibitem[{\citenamefont{{Kirkman} et~al.}(2003)\citenamefont{{Kirkman},
  {Tytler}, {Suzuki}, {O'Meara}, and {Lubin}}}]{kirkman}
\bibinfo{author}{\bibfnamefont{D.}~\bibnamefont{{Kirkman}}},
  \bibinfo{author}{\bibfnamefont{D.}~\bibnamefont{{Tytler}}},
  \bibinfo{author}{\bibfnamefont{N.}~\bibnamefont{{Suzuki}}},
  \bibinfo{author}{\bibfnamefont{J.~M.} \bibnamefont{{O'Meara}}},
  \bibnamefont{and} \bibinfo{author}{\bibfnamefont{D.}~\bibnamefont{{Lubin}}},
  \bibinfo{journal}{Astrophys.J.Suppl.Ser.} \textbf{\bibinfo{volume}{149}},
  \bibinfo{pages}{1} (\bibinfo{year}{2003}).

\bibitem[{\citenamefont{{Burles} and {Tytler}}(1998{\natexlab{a}})}]{burles1}
\bibinfo{author}{\bibfnamefont{S.}~\bibnamefont{{Burles}}} \bibnamefont{and}
  \bibinfo{author}{\bibfnamefont{D.}~\bibnamefont{{Tytler}}},
  \bibinfo{journal}{Astrophys.J.} \textbf{\bibinfo{volume}{499}},
  \bibinfo{pages}{699} (\bibinfo{year}{1998}{\natexlab{a}}).

\bibitem[{\citenamefont{{Burles} and {Tytler}}(1998{\natexlab{b}})}]{burles2}
\bibinfo{author}{\bibfnamefont{S.}~\bibnamefont{{Burles}}} \bibnamefont{and}
  \bibinfo{author}{\bibfnamefont{D.}~\bibnamefont{{Tytler}}},
  \bibinfo{journal}{Astrophys.J.} \textbf{\bibinfo{volume}{507}},
  \bibinfo{pages}{732} (\bibinfo{year}{1998}{\natexlab{b}}).

\bibitem[{\citenamefont{{Crighton} et~al.}(2004)\citenamefont{{Crighton},
  {Webb}, {Ortiz-Gil}, and {Fern{\' a}ndez-Soto}}}]{Crighton04}
\bibinfo{author}{\bibfnamefont{N.~H.~M.} \bibnamefont{{Crighton}}},
  \bibinfo{author}{\bibfnamefont{J.~K.} \bibnamefont{{Webb}}},
  \bibinfo{author}{\bibfnamefont{A.}~\bibnamefont{{Ortiz-Gil}}},
  \bibnamefont{and} \bibinfo{author}{\bibfnamefont{A.}~\bibnamefont{{Fern{\'
  a}ndez-Soto}}}, \bibinfo{journal}{Mon.Not.R.Astron.Soc.}
  \textbf{\bibinfo{volume}{355}}, \bibinfo{pages}{1042} (\bibinfo{year}{2004}).

\bibitem[{\citenamefont{{O'Meara} et~al.}(2006)\citenamefont{{O'Meara},
  {Burles}, {Prochaska}, {Prochter}, {Bernstein}, and {Burgess}}}]{omeara06}
\bibinfo{author}{\bibfnamefont{J.~M.} \bibnamefont{{O'Meara}}},
  \bibinfo{author}{\bibfnamefont{S.}~\bibnamefont{{Burles}}},
  \bibinfo{author}{\bibfnamefont{J.~X.} \bibnamefont{{Prochaska}}},
  \bibinfo{author}{\bibfnamefont{G.~E.} \bibnamefont{{Prochter}}},
  \bibinfo{author}{\bibfnamefont{R.~A.} \bibnamefont{{Bernstein}}},
  \bibnamefont{and} \bibinfo{author}{\bibfnamefont{K.~M.}
  \bibnamefont{{Burgess}}}, \bibinfo{journal}{Astrophys.J.Lett.}
  \textbf{\bibinfo{volume}{649}}, \bibinfo{pages}{L61} (\bibinfo{year}{2006}),
  \eprint{astro-ph/0608302}.

\bibitem[{\citenamefont{{Oliveira} et~al.}(2006)\citenamefont{{Oliveira},
  {Moos}, {Chayer}, and {Kruk}}}]{oliveira06}
\bibinfo{author}{\bibfnamefont{C.~M.} \bibnamefont{{Oliveira}}},
  \bibinfo{author}{\bibfnamefont{H.~W.} \bibnamefont{{Moos}}},
  \bibinfo{author}{\bibfnamefont{P.}~\bibnamefont{{Chayer}}}, \bibnamefont{and}
  \bibinfo{author}{\bibfnamefont{J.~W.} \bibnamefont{{Kruk}}},
  \bibinfo{journal}{Astrophys.J.} \textbf{\bibinfo{volume}{642}},
  \bibinfo{pages}{283} (\bibinfo{year}{2006}), \eprint{astro-ph/0601114}.

\bibitem[{\citenamefont{{Ryan} et~al.}(2000)\citenamefont{{Ryan}, {Beers},
  {Olive}, {Fields}, and {Norris}}}]{ryan}
\bibinfo{author}{\bibfnamefont{S.}~\bibnamefont{{Ryan}}},
  \bibinfo{author}{\bibfnamefont{T.}~\bibnamefont{{Beers}}},
  \bibinfo{author}{\bibfnamefont{K.}~\bibnamefont{{Olive}}},
  \bibinfo{author}{\bibfnamefont{B.~D.} \bibnamefont{{Fields}}},
  \bibnamefont{and} \bibinfo{author}{\bibfnamefont{J.~E.}
  \bibnamefont{{Norris}}}, \bibinfo{journal}{Astrophys.J.}
  \textbf{\bibinfo{volume}{530}}, \bibinfo{pages}{L57} (\bibinfo{year}{2000}).

\bibitem[{\citenamefont{{Bonifacio} et~al.}(1997)\citenamefont{{Bonifacio},
  {Molaro}, and {Pasquini}}}]{bonifacio1}
\bibinfo{author}{\bibfnamefont{P.}~\bibnamefont{{Bonifacio}}},
  \bibinfo{author}{\bibfnamefont{P.}~\bibnamefont{{Molaro}}}, \bibnamefont{and}
  \bibinfo{author}{\bibfnamefont{L.}~\bibnamefont{{Pasquini}}},
  \bibinfo{journal}{Mon.Not.R.Astron.Soc.} \textbf{\bibinfo{volume}{292}},
  \bibinfo{pages}{L1} (\bibinfo{year}{1997}).

\bibitem[{\citenamefont{{Bonifacio} and {Molaro}}(1997)}]{bonifacio2}
\bibinfo{author}{\bibfnamefont{P.}~\bibnamefont{{Bonifacio}}} \bibnamefont{and}
  \bibinfo{author}{\bibfnamefont{P.}~\bibnamefont{{Molaro}}},
  \bibinfo{journal}{Mon.Not.R.Astron.Soc.} \textbf{\bibinfo{volume}{285}},
  \bibinfo{pages}{847} (\bibinfo{year}{1997}).

\bibitem[{\citenamefont{{Bonifacio et al}}(2002)}]{bonifacio3}
\bibinfo{author}{\bibnamefont{{Bonifacio et al}}}, \bibinfo{journal}{Astronomy
  and Astrophysics} \textbf{\bibinfo{volume}{390}}, \bibinfo{pages}{91}
  (\bibinfo{year}{2002}).

\bibitem[{\citenamefont{{Asplund} et~al.}(2006)\citenamefont{{Asplund},
  {Lambert}, {Nissen}, {Primas}, and {Smith}}}]{Asplund05}
\bibinfo{author}{\bibfnamefont{M.}~\bibnamefont{{Asplund}}},
  \bibinfo{author}{\bibfnamefont{D.~L.} \bibnamefont{{Lambert}}},
  \bibinfo{author}{\bibfnamefont{P.~E.} \bibnamefont{{Nissen}}},
  \bibinfo{author}{\bibfnamefont{F.}~\bibnamefont{{Primas}}}, \bibnamefont{and}
  \bibinfo{author}{\bibfnamefont{V.~V.} \bibnamefont{{Smith}}},
  \bibinfo{journal}{Astrophys.J.} \textbf{\bibinfo{volume}{644}},
  \bibinfo{pages}{229} (\bibinfo{year}{2006}), \eprint{astro-ph/0510636}.

\bibitem[{\citenamefont{{Boesgaard} et~al.}(2005)\citenamefont{{Boesgaard},
  {Novicki}, and {Stephens}}}]{BNS05}
\bibinfo{author}{\bibfnamefont{A.~M.} \bibnamefont{{Boesgaard}}},
  \bibinfo{author}{\bibfnamefont{M.~C.} \bibnamefont{{Novicki}}},
  \bibnamefont{and}
  \bibinfo{author}{\bibfnamefont{A.}~\bibnamefont{{Stephens}}}, in
  \emph{\bibinfo{booktitle}{Proceedings of IAU Symposium No. 228: "From Lithium
  to Uranium: Elemental Tracers of Early Cosmic Evolution"}}, edited by
  \bibinfo{editor}{\bibfnamefont{V.}~\bibnamefont{{Hill}}},
  \bibinfo{editor}{\bibfnamefont{P.}~\bibnamefont{{Francois}}},
  \bibnamefont{and} \bibinfo{editor}{\bibfnamefont{F.}~\bibnamefont{{Primas}}}
  (\bibinfo{publisher}{Cambridge University Press}, \bibinfo{year}{2005}),
  p.~\bibinfo{pages}{29}.

\bibitem[{\citenamefont{{Bonifacio} et~al.}(2007)\citenamefont{{Bonifacio},
  {Molaro}, {Sivarani}, {Cayrel}, {Spite}, {Spite}, {Plez}, {Andersen},
  {Barbuy}, {Beers} et~al.}}]{bonifacio07}
\bibinfo{author}{\bibfnamefont{P.}~\bibnamefont{{Bonifacio}}},
  \bibinfo{author}{\bibfnamefont{P.}~\bibnamefont{{Molaro}}},
  \bibinfo{author}{\bibfnamefont{T.}~\bibnamefont{{Sivarani}}},
  \bibinfo{author}{\bibfnamefont{R.}~\bibnamefont{{Cayrel}}},
  \bibinfo{author}{\bibfnamefont{M.}~\bibnamefont{{Spite}}},
  \bibinfo{author}{\bibfnamefont{F.}~\bibnamefont{{Spite}}},
  \bibinfo{author}{\bibfnamefont{B.}~\bibnamefont{{Plez}}},
  \bibinfo{author}{\bibfnamefont{J.}~\bibnamefont{{Andersen}}},
  \bibinfo{author}{\bibfnamefont{B.}~\bibnamefont{{Barbuy}}},
  \bibinfo{author}{\bibfnamefont{T.~C.} \bibnamefont{{Beers}}},
  \bibnamefont{et~al.}, \bibinfo{journal}{Astron. and Astrophys.}
  \textbf{\bibinfo{volume}{462}}, \bibinfo{pages}{851} (\bibinfo{year}{2007}),
  \eprint{arXiv:astro-ph/0610245}.

\bibitem[{\citenamefont{{Peimbert} et~al.}(2007)\citenamefont{{Peimbert},
  {Luridiana}, and {Peimbert}}}]{PL07}
\bibinfo{author}{\bibfnamefont{M.}~\bibnamefont{{Peimbert}}},
  \bibinfo{author}{\bibfnamefont{V.}~\bibnamefont{{Luridiana}}},
  \bibnamefont{and}
  \bibinfo{author}{\bibfnamefont{A.}~\bibnamefont{{Peimbert}}},
  \bibinfo{journal}{\apj} \textbf{\bibinfo{volume}{666}}, \bibinfo{pages}{636}
  (\bibinfo{year}{2007}), \eprint{arXiv:astro-ph/0701580}.

\bibitem[{\citenamefont{{Izotov} et~al.}(2007)\citenamefont{{Izotov}, {Thuan},
  and {Stasi{\'n}ska}}}]{izotov07}
\bibinfo{author}{\bibfnamefont{Y.~I.} \bibnamefont{{Izotov}}},
  \bibinfo{author}{\bibfnamefont{T.~X.} \bibnamefont{{Thuan}}},
  \bibnamefont{and}
  \bibinfo{author}{\bibfnamefont{G.}~\bibnamefont{{Stasi{\'n}ska}}},
  \bibinfo{journal}{\apj} \textbf{\bibinfo{volume}{662}}, \bibinfo{pages}{15}
  (\bibinfo{year}{2007}), \eprint{arXiv:astro-ph/0702072}.

\bibitem[{\citenamefont{{Yao et al.}}(2006)}]{PDGBook}
\bibinfo{author}{\bibnamefont{{Yao et al.}}}, \bibinfo{journal}{{Journal of
  Physics G}} \textbf{\bibinfo{volume}{33}}, \bibinfo{pages}{1}
  (\bibinfo{year}{2006}), \urlprefix\url{http://pdg.lbl.gov}.

\bibitem[{\citenamefont{{Olive} and {Skillman}}(2004)}]{olive07}
\bibinfo{author}{\bibfnamefont{K.~A.} \bibnamefont{{Olive}}} \bibnamefont{and}
  \bibinfo{author}{\bibfnamefont{E.~D.} \bibnamefont{{Skillman}}},
  \bibinfo{journal}{\apj} \textbf{\bibinfo{volume}{617}}, \bibinfo{pages}{29}
  (\bibinfo{year}{2004}), \eprint{arXiv:astro-ph/0405588}.

\bibitem[{\citenamefont{Kawano}(1992)}]{Kawano92}
\bibinfo{author}{\bibfnamefont{L.}~\bibnamefont{Kawano}}
  (\bibinfo{year}{1992}), \bibinfo{note}{fERMILAB-PUB-92-004-A}.

\bibitem[{\citenamefont{{Dicus} et~al.}(1982)\citenamefont{{Dicus}, {Kolb},
  {Gleeson}, {Sudarshan}, {Teplitz}, and {Turner}}}]{Dicus82}
\bibinfo{author}{\bibfnamefont{D.~A.} \bibnamefont{{Dicus}}},
  \bibinfo{author}{\bibfnamefont{E.~W.} \bibnamefont{{Kolb}}},
  \bibinfo{author}{\bibfnamefont{A.~M.} \bibnamefont{{Gleeson}}},
  \bibinfo{author}{\bibfnamefont{E.~C.~G.} \bibnamefont{{Sudarshan}}},
  \bibinfo{author}{\bibfnamefont{V.~L.} \bibnamefont{{Teplitz}}},
  \bibnamefont{and} \bibinfo{author}{\bibfnamefont{M.~S.}
  \bibnamefont{{Turner}}}, \bibinfo{journal}{\prd}
  \textbf{\bibinfo{volume}{26}}, \bibinfo{pages}{2694} (\bibinfo{year}{1982}).

\bibitem[{\citenamefont{{Beane} and {Savage}}(2003{\natexlab{a}})}]{BS03b}
\bibinfo{author}{\bibfnamefont{S.~R.} \bibnamefont{{Beane}}} \bibnamefont{and}
  \bibinfo{author}{\bibfnamefont{M.~J.} \bibnamefont{{Savage}}},
  \bibinfo{journal}{Nuclear Physics A} \textbf{\bibinfo{volume}{717}},
  \bibinfo{pages}{91} (\bibinfo{year}{2003}{\natexlab{a}}).

\bibitem[{\citenamefont{{Reid}}(1968)}]{reid68}
\bibinfo{author}{\bibfnamefont{R.~V.} \bibnamefont{{Reid}},
  \bibfnamefont{Jr.}}, \bibinfo{journal}{Annals of Physics}
  \textbf{\bibinfo{volume}{50}}, \bibinfo{pages}{411} (\bibinfo{year}{1968}).

\bibitem[{\citenamefont{{Epelbaum} et~al.}(2003)\citenamefont{{Epelbaum},
  {Mei{\ss}ner}, and {Gl{\" o}ckle}}}]{EMG03}
\bibinfo{author}{\bibfnamefont{E.}~\bibnamefont{{Epelbaum}}},
  \bibinfo{author}{\bibfnamefont{U.}~\bibnamefont{{Mei{\ss}ner}}},
  \bibnamefont{and} \bibinfo{author}{\bibfnamefont{W.}~\bibnamefont{{Gl{\"
  o}ckle}}}, \bibinfo{journal}{Nuclear Physics A}
  \textbf{\bibinfo{volume}{714}}, \bibinfo{pages}{535} (\bibinfo{year}{2003}).

\bibitem[{\citenamefont{{Richard} et~al.}(2005)\citenamefont{{Richard},
  {Michaud}, and {Richer}}}]{richard05}
\bibinfo{author}{\bibfnamefont{O.}~\bibnamefont{{Richard}}},
  \bibinfo{author}{\bibfnamefont{G.}~\bibnamefont{{Michaud}}},
  \bibnamefont{and} \bibinfo{author}{\bibfnamefont{J.}~\bibnamefont{{Richer}}},
  \bibinfo{journal}{Astrophys.J.} \textbf{\bibinfo{volume}{619}},
  \bibinfo{pages}{538} (\bibinfo{year}{2005}).

\bibitem[{\citenamefont{{Mel{\' e}ndez} and {Ram{\'{\i}}rez}}(2004)}]{MR04}
\bibinfo{author}{\bibfnamefont{J.}~\bibnamefont{{Mel{\' e}ndez}}}
  \bibnamefont{and}
  \bibinfo{author}{\bibfnamefont{I.}~\bibnamefont{{Ram{\'{\i}}rez}}},
  \bibinfo{journal}{Astrophys.J.Lett.} \textbf{\bibinfo{volume}{615}},
  \bibinfo{pages}{L33} (\bibinfo{year}{2004}).

\bibitem[{\citenamefont{{Prodanovi{\'c}} and {Fields}}(2007)}]{PF07}
\bibinfo{author}{\bibfnamefont{T.}~\bibnamefont{{Prodanovi{\'c}}}}
  \bibnamefont{and} \bibinfo{author}{\bibfnamefont{B.~D.}
  \bibnamefont{{Fields}}}, \bibinfo{journal}{\prd}
  \textbf{\bibinfo{volume}{76}}, \bibinfo{pages}{083003}
  (\bibinfo{year}{2007}), \eprint{arXiv:0709.3300}.

\bibitem[{\citenamefont{{Battye} et~al.}(2001)\citenamefont{{Battye},
  {Crittenden}, and {Weller}}}]{BCW01}
\bibinfo{author}{\bibfnamefont{R.~A.} \bibnamefont{{Battye}}},
  \bibinfo{author}{\bibfnamefont{R.}~\bibnamefont{{Crittenden}}},
  \bibnamefont{and} \bibinfo{author}{\bibfnamefont{J.}~\bibnamefont{{Weller}}},
  \bibinfo{journal}{\prd} \textbf{\bibinfo{volume}{63}},
  \bibinfo{pages}{043505} (\bibinfo{year}{2001}),
  \eprint{arXiv:astro-ph/0008265}.

\bibitem[{\citenamefont{{Landau} et~al.}(2001)\citenamefont{{Landau}, {Harari},
  and {Zaldarriaga}}}]{LHZ01}
\bibinfo{author}{\bibfnamefont{S.~J.} \bibnamefont{{Landau}}},
  \bibinfo{author}{\bibfnamefont{D.~D.} \bibnamefont{{Harari}}},
  \bibnamefont{and}
  \bibinfo{author}{\bibfnamefont{M.}~\bibnamefont{{Zaldarriaga}}},
  \bibinfo{journal}{\prd} \textbf{\bibinfo{volume}{63}},
  \bibinfo{pages}{083505} (\bibinfo{year}{2001}),
  \eprint{arXiv:astro-ph/0010415}.

\bibitem[{\citenamefont{{Avelino} et~al.}(2000)\citenamefont{{Avelino},
  {Martins}, {Rocha}, and {Viana}}}]{AV00}
\bibinfo{author}{\bibfnamefont{P.~P.} \bibnamefont{{Avelino}}},
  \bibinfo{author}{\bibfnamefont{C.~J.~A.~P.} \bibnamefont{{Martins}}},
  \bibinfo{author}{\bibfnamefont{G.}~\bibnamefont{{Rocha}}}, \bibnamefont{and}
  \bibinfo{author}{\bibfnamefont{P.}~\bibnamefont{{Viana}}},
  \bibinfo{journal}{\prd} \textbf{\bibinfo{volume}{62}},
  \bibinfo{pages}{123508} (\bibinfo{year}{2000}),
  \eprint{arXiv:astro-ph/0008446}.

\bibitem[{\citenamefont{Readhead et~al.}(2004)}]{CBI04}
\bibinfo{author}{\bibfnamefont{A.~C.~S.} \bibnamefont{Readhead}}
  \bibnamefont{et~al.}, \bibinfo{journal}{Astrophys. J.}
  \textbf{\bibinfo{volume}{609}}, \bibinfo{pages}{498} (\bibinfo{year}{2004}),
  \eprint{astro-ph/0402359}.

\bibitem[{\citenamefont{Kuo et~al.}(2004)}]{ACBAR02}
\bibinfo{author}{\bibfnamefont{C.}~\bibnamefont{Kuo}} \bibnamefont{et~al.}
  (\bibinfo{collaboration}{ACBAR}), \bibinfo{journal}{Astrophys. J.}
  \textbf{\bibinfo{volume}{600}}, \bibinfo{pages}{32} (\bibinfo{year}{2004}),
  \eprint{astro-ph/0212289}.

\bibitem[{\citenamefont{Piacentini et~al.}(2006)}]{BOOM05_polar}
\bibinfo{author}{\bibfnamefont{F.}~\bibnamefont{Piacentini}}
  \bibnamefont{et~al.}, \bibinfo{journal}{Astrophys.J.}
  \textbf{\bibinfo{volume}{647}}, \bibinfo{pages}{833} (\bibinfo{year}{2006}),
  \eprint{astro-ph/0507507}.

\bibitem[{\citenamefont{Jones et~al.}(2006)}]{BOOM05_temp}
\bibinfo{author}{\bibfnamefont{W.~C.} \bibnamefont{Jones}}
  \bibnamefont{et~al.}, \bibinfo{journal}{Astrophys.J.}
  \textbf{\bibinfo{volume}{647}}, \bibinfo{pages}{823} (\bibinfo{year}{2006}),
  \eprint{astro-ph/0507494}.

\bibitem[{\citenamefont{Cole et~al.}(2005)}]{2dF05}
\bibinfo{author}{\bibfnamefont{S.}~\bibnamefont{Cole}} \bibnamefont{et~al.}
  (\bibinfo{collaboration}{The 2dFGRS}), \bibinfo{journal}{Mon. Not. Roy.
  Astron. Soc.} \textbf{\bibinfo{volume}{362}}, \bibinfo{pages}{505}
  (\bibinfo{year}{2005}), \eprint{astro-ph/0501174}.

\bibitem[{\citenamefont{{Lewis} and {Bridle}}(2002)}]{LB02}
\bibinfo{author}{\bibfnamefont{A.}~\bibnamefont{{Lewis}}} \bibnamefont{and}
  \bibinfo{author}{\bibfnamefont{S.}~\bibnamefont{{Bridle}}},
  \bibinfo{journal}{\prd} \textbf{\bibinfo{volume}{66}},
  \bibinfo{pages}{103511} (\bibinfo{year}{2002}), \eprint{astro-ph/0205436}.

\bibitem[{\citenamefont{{Lewis} et~al.}(2000)\citenamefont{{Lewis},
  {Challinor}, and {Lasenby}}}]{LCL00}
\bibinfo{author}{\bibfnamefont{A.}~\bibnamefont{{Lewis}}},
  \bibinfo{author}{\bibfnamefont{A.}~\bibnamefont{{Challinor}}},
  \bibnamefont{and}
  \bibinfo{author}{\bibfnamefont{A.}~\bibnamefont{{Lasenby}}},
  \bibinfo{journal}{Astrophys.J.} \textbf{\bibinfo{volume}{538}},
  \bibinfo{pages}{473} (\bibinfo{year}{2000}), \eprint{astro-ph/9911177}.

\bibitem[{\citenamefont{{Seager} et~al.}(1999)\citenamefont{{Seager},
  {Sasselov}, and {Scott}}}]{recfast}
\bibinfo{author}{\bibfnamefont{S.}~\bibnamefont{{Seager}}},
  \bibinfo{author}{\bibfnamefont{D.~D.} \bibnamefont{{Sasselov}}},
  \bibnamefont{and} \bibinfo{author}{\bibfnamefont{D.}~\bibnamefont{{Scott}}},
  \bibinfo{journal}{Astrophys.J.Lett.} \textbf{\bibinfo{volume}{523}},
  \bibinfo{pages}{L1} (\bibinfo{year}{1999}), \eprint{astro-ph/9909275}.

\bibitem[{\citenamefont{Raftery and Lewis}(1992)}]{Raftery&Lewis}
\bibinfo{author}{\bibfnamefont{A.~E.} \bibnamefont{Raftery}} \bibnamefont{and}
  \bibinfo{author}{\bibfnamefont{S.~M.} \bibnamefont{Lewis}}, in
  \emph{\bibinfo{booktitle}{Bayesian Statistics}}, edited by
  \bibinfo{editor}{\bibfnamefont{J.~M.} \bibnamefont{Bernado}}
  (\bibinfo{publisher}{OUP}, \bibinfo{year}{1992}), p. \bibinfo{pages}{765}.

\bibitem[{\citenamefont{{Christiansen}
  et~al.}(1991)\citenamefont{{Christiansen}, {Epele}, {Fanchiotti}, and
  {Garc{\'{\i}}a Canal}}}]{Epele91b}
\bibinfo{author}{\bibfnamefont{H.~R.} \bibnamefont{{Christiansen}}},
  \bibinfo{author}{\bibfnamefont{L.~N.} \bibnamefont{{Epele}}},
  \bibinfo{author}{\bibfnamefont{H.}~\bibnamefont{{Fanchiotti}}},
  \bibnamefont{and} \bibinfo{author}{\bibfnamefont{C.~A.}
  \bibnamefont{{Garc{\'{\i}}a Canal}}}, \bibinfo{journal}{Physics Letters B}
  \textbf{\bibinfo{volume}{267}}, \bibinfo{pages}{164} (\bibinfo{year}{1991}).

\bibitem[{\citenamefont{{Kawano}}(1988)}]{Kawano88}
\bibinfo{author}{\bibfnamefont{L.}~\bibnamefont{{Kawano}}}
  (\bibinfo{year}{1988}), \bibinfo{note}{fERMILAB-PUB-88-034-A}.

\bibitem[{\citenamefont{{Dixit} and {Sher}}(1988)}]{dixit88}
\bibinfo{author}{\bibfnamefont{V.~V.} \bibnamefont{{Dixit}}} \bibnamefont{and}
  \bibinfo{author}{\bibfnamefont{M.}~\bibnamefont{{Sher}}},
  \bibinfo{journal}{\prd} \textbf{\bibinfo{volume}{37}}, \bibinfo{pages}{1097}
  (\bibinfo{year}{1988}).

\bibitem[{\citenamefont{{Beane} and {Savage}}(2003{\natexlab{b}})}]{BS03}
\bibinfo{author}{\bibfnamefont{S.~R.} \bibnamefont{{Beane}}} \bibnamefont{and}
  \bibinfo{author}{\bibfnamefont{M.~J.} \bibnamefont{{Savage}}},
  \bibinfo{journal}{Nuclear Physics A} \textbf{\bibinfo{volume}{713}},
  \bibinfo{pages}{148} (\bibinfo{year}{2003}{\natexlab{b}}).

\bibitem[{\citenamefont{{Peebles}}(1968)}]{Peebles68}
\bibinfo{author}{\bibfnamefont{P.~J.~E.} \bibnamefont{{Peebles}}},
  \bibinfo{journal}{Astrophys.J.} \textbf{\bibinfo{volume}{153}},
  \bibinfo{pages}{1} (\bibinfo{year}{1968}).

\end{thebibliography}
\bibliographystyle{apsrev}

\end{document}